\documentclass[%
 reprint,
 superscriptaddress,
 amsmath,amssymb,
 prb,
]{revtex4-2}

\UseRawInputEncoding

\usepackage{graphicx}
\usepackage{dcolumn}
\usepackage{bm}
\usepackage{graphicx}
\usepackage{color}
\usepackage[colorlinks, linkcolor=blue, anchorcolor=blue, citecolor=blue, urlcolor=blue]{hyperref}
\usepackage{amsmath}
\usepackage{multirow}
\usepackage{tabularx}
\usepackage{longtable}
\usepackage{rotating}

\newcommand{\nmao}{NdMgAl$_{11}$O$_{19}$}
\newcommand{\pmao}{PrMgAl$_{11}$O$_{19}$}
\newcommand{\cmao}{CeMgAl$_{11}$O$_{19}$}

\newcommand{\GGG}{Gd$_3$Ga$_{5}$O$_{12}$}

\emergencystretch=\maxdimen
\begin{document}


\title{Giant magnetocaloric effect at low fields in triangular-lattice \nmao}


\author{Yantao Cao}
\affiliation{Beijing National Laboratory for Condensed Matter Physics, Institute of Physics, Chinese Academy of Sciences, Beijing 100190, China}
\affiliation{Songshan Lake Materials Laboratory, Dongguan 523808, China}
\affiliation{School of Physical Sciences, University of Chinese Academy of Sciences, Beijing 100049, China}
\author{He Sun}
\affiliation{Beijing National Laboratory for Condensed Matter Physics, Institute of Physics, Chinese Academy of Sciences, Beijing 100190, China}
\author{Zhendong Fu}
\affiliation{Songshan Lake Materials Laboratory, Dongguan 523808, China}
\author{Zhaoming Tian}
\affiliation{School of Physics and Wuhan National High Magnetic Field Center,
Huazhong University of Science and Technology, Wuhan 430074, China}
\author{Huiqian Luo}
\affiliation{Beijing National Laboratory for Condensed Matter Physics, Institute of Physics, Chinese Academy of Sciences, Beijing 100190, China}
\author{Junsen Xiang}
\affiliation{Beijing National Laboratory for Condensed Matter Physics, Institute of Physics, Chinese Academy of Sciences, Beijing 100190, China}
\author{Peijie Sun}
\email{pjsun@iphy.ac.cn}
\affiliation{Beijing National Laboratory for Condensed Matter Physics, Institute of Physics, Chinese Academy of Sciences, Beijing 100190, China}
\author{Jinkui Zhao}
\email{jkzhao@gbu.edu.cn}
\affiliation{School of Physical Sciences, Great Bay University, Dongguan 523808, China}
\author{Hanjie Guo}
\email{hjguo@sslab.org.cn}
\affiliation{Songshan Lake Materials Laboratory, Dongguan 523808, China}



\date{\today}

\begin{abstract}
  Magnetic refrigeration in the sub-Kelvin regime requires refrigerant materials to retain a large magnetic entropy at low temperatures by suppressing magnetic ordering. Quantum spin liquids (QSLs), which evade long-range magnetic ordering while retaining strong quantum fluctuations to the lowest temperatures, therefore provide a promising platform for realizing high-performance magnetic refrigerants. Here, we investigate the magnetic ground state and the magnetocaloric effect of the hexaaluminate, \nmao, in which the Nd$^{3+}$ ions form a network of triangular lattices. Magnetic susceptibility and specific heat measurements indicate a magnetically dynamic state down to 50~mK, consistent with a QSL state. Specific heat measurements further reveal substantial magnetic entropy retained below 50~mK. Quasi-adiabatic demagnetization measurements demonstrate a superior cooling performance of \nmao, which can be cooled to 113~mK from 1.9~K by only a small magnetic field change of 2~T. The outstanding refrigeration performance is attributed to the persistent spin fluctuations associated with the QSL-like ground state, together with a large effective \textit{g} factor and the smallness of the exchange interactions along the easy-axis direction. This study demonstrates that frustration, combined with strong spin-orbit coupling and crystal-electric-field effect in the rare earth magnets provides a promising design principle for next-generation cryogenic magnetic refrigerants.
\end{abstract}


\maketitle

\section{INTRODUCTION}

Low temperatures extending down to the sub-Kelvin regime play a crucial role in quantum information processing \cite{gaita2019molecular,petit2020universal}, high-sensitivity sensing technologies \cite{kittel1980temperature,kittel1981refrigeration}, and the exploration of emergent quantum phenomena in condensed matter physics \cite{RevModPhys.86.563,RevModPhys.89.025003,RevModPhys.83.1193}. Among the various approaches for reaching such temperatures, adiabatic demagnetization refrigeration (ADR) has attracted considerable attention owing to its compact design \cite{shirron2014optimization}, operational simplicity, and independence from scarce $^3$He resources. Unlike dilution refrigeration, ADR relies on the magnetocaloric effect (MCE), where the entropy of the magnetic sub-system increases as the magnetic field is adiabatically reduced, leading to a substantial decrease in temperature.
A central challenge in the development of sub-Kelvin ADR materials is to suppress long-range magnetic ordering, thereby preserving a large amount of entropy down to the lowest temperatures \cite{tokiwa2021frustrated,PhysRevB.67.104421,sosin2005magnetocaloric,shu2026giant,treu2025utilizing,song2025realization,PhysRevB.107.104402,lin2025quantum,brasiliano2020ybgg,xiang2024giant}. This is achieved in conventional paramagnetic salts by separating the magnetic ions with nonmagnetic ligands, such as water molecules, to minimize magnetic interactions ($J$) between neighboring spins \cite{kittel1980temperature,PhysRev.148.509,fritz1975magnetic}. Nevertheless, magnetic order can still occur at temperatures on the order of $J$, leading to a rapid loss of magnetic entropy and ultimately limiting the minimum achievable temperature. An alternative strategy is provided by frustrated magnets where a long-range order is suppressed despite appreciable exchange interactions. In this context, quantum spin liquids, in which the spins resist ordering even down to absolute zero, have been widely considered as potential platforms for ADR \cite{balents2010spin,savary2016quantum,broholm2020quantum,PhysRevB.109.155129,liu2022quantum}.

Despite extensive efforts devoted to frustrated magnets and paramagnetic salts for ADR applications, \GGG\ (GGG) remains the benchmark refrigerant for sub-Kelvin cooling. This is primarily attributed to its low magnetic ordering temperature and large magnetic entropy reservoir arising from strong geometric frustration \cite{daudin1982thermodynamic,dai1988magnetothermal}. However, for GGG and other Gd-based systems, the large magnetocaloric response is typically achieved only under relatively high magnetic fields, restricting their applicability in compact refrigeration devices based on permanent magnets. To overcome this limitation, it is desirable to identify frustrated magnetic systems that simultaneously preserve a large magnetic entropy and exhibit an enhanced magnetocaloric response under low magnetic fields. Rare-earth systems with unquenched orbital moments offer an attractive platform in this regard, as the interplay between spin-orbit coupling, crystal-electric-field (CEF) splitting, and exchange interactions can generate highly anisotropic effective spin states and suppress magnetic ordering.

In this work, we investigate the ADR performance of a Nd-based quantum-spin-liquid candidate, the rare-earth hexaaluminate \nmao, in which Nd$^{3+}$ ions form a two-dimensional triangular-lattice network.
Recent studies on this family of compounds with other rare-earth elements have indicated the existence of potential QSL states with Ising-dominant anisotropy \cite{ashtar2019,Bu2022,cao2024synthesis,cao2025u,Li2024,Ma2024}.
In addition, this compound can be grown as centimeter-sized high-quality single crystals and exhibits excellent chemical stability, providing an ideal material platform for potential practical engineering applications.
Here, we demonstrate the absence of long-range magnetic ordering down to 50 mK for \nmao, despite strong interactions among the Nd$^{3+}$ ions. Detailed specific heat measurements further reveal that a substantial amount of magnetic entropy is retained well below 50~mK, consistent with a highly dynamic magnetic ground state. Together with a large effective \textit{g} factor along the \textit{c} axis, these characteristics suggest a potentially strong magnetocaloric effect. This is corroborated by direct adiabatic demagnetization measurements, which show a pronounced cooling effect, with the temperature decreasing from 1.9~K to 113 mK under a small field change of only 2 T. Such a performance outperforms the commercial benchmark \GGG, particularly in the low-field regime.

\section{EXPERIMENTAL SECTION}

Polycrystalline samples were synthesized using a standard solid-state reaction technique. Raw materials of Nd$_2$O$_3$ (99.99\%), MgO (99.99\%), and Al$_2$O$_3$ (99.99\%) were dried at 900$^{\circ}$C over night prior to reaction to avoid moisture contamination. Then, stoichiometric amounts of the raw materials were mixed and ground thoroughly, pressed into pellets and sintered at 1500$^{\circ}$C $\sim$ 1600$^{\circ}$C with several intermediate grindings. The powders were later made into rods (13 cm in length and 7 mm in diameter) through hydrostatic pressure at 70 MPa and these rods were then sintered at 1500$^{\circ}$C for 2 hours with an intermediate grinding. Subsequent single-crystal growth was conducted in an optical floating zone furnace in pure argon atmosphere at 9 bar. After the floating-zone growth, the several-centimeter-sized crystal was annealed at 1000$^{\circ}$C in a flowing oxygen atmosphere for 24 h and then slowly cooled down to room temperature in order to avoid any possible oxygen vacancy.

Single-crystal X-ray diffraction (XRD) measurements were performed on a XtaLAB Synergy diffractometer (Rigaku) at room temperature using the Mo-$K_\alpha$ radiation. The experimental conditions are tabulated in the Supporting Information (SI) Tab. S1. Part of the single crystal was crushed into powder for powder XRD measurements performed on a MiniFlex diffractometer (Rigaku) with the Cu-$K_\alpha$ radiation. JANA2020 \cite{petvrivcek2023jana2020} and FULLPROF \cite{rodriguez1993recent} packages were used for crystal structure refinement.

Specific heat measurements were carried out on a physical property measurement system (PPMS, Quantum Design) equipped with a dilution refrigerator insert. DC magnetic susceptibility was measured over the temperature range of 2-300 K using the vibrating sample magnetometer (VSM) option of the PPMS. AC magnetic susceptibility was measured using the ACMS-II and ACDR options, with an activation field of 1-3 Oe in amplitude.

Quasi-adiabatic demagnetization cooling measurements were carried out using an adiabatic magneto-calorimetry system \cite{xiang2024giant}. Approximately 1.6 g of the single crystals were coaligned with their crystallographic \textit{c} axes parallel to the applied magnetic field direction. The sample temperature was recorded while the magnetic field was gradually decreased from a specific initial field at a fixed initial temperature.

\section{RESULTS AND DISCUSSION}

\subsection{Crystal Structure}

\begin{figure}
	\centering
	\includegraphics[width=1\columnwidth]{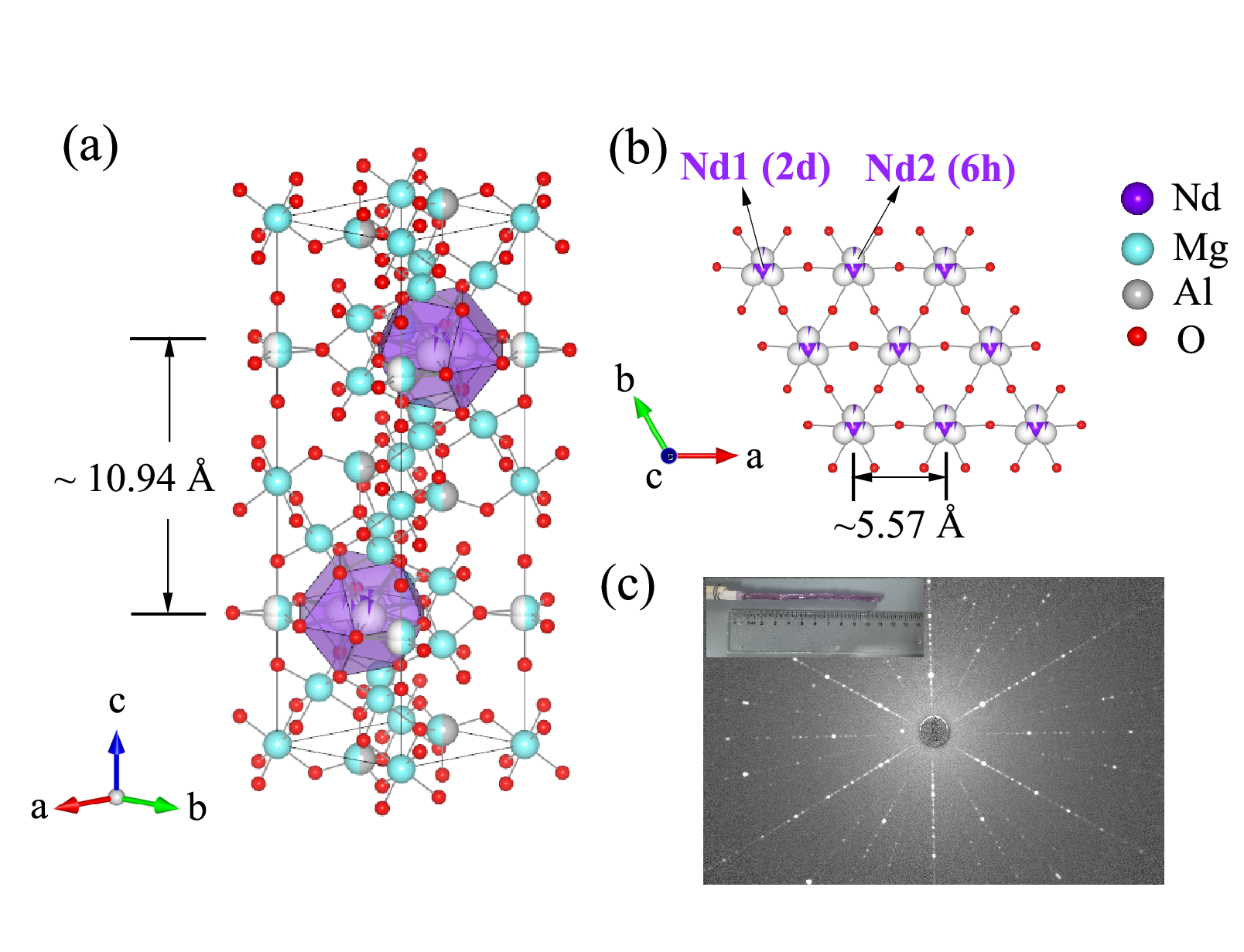}
	\caption{Crystal structure. (a) Crystal structure of \nmao\ determined by single-crystal XRD refinements. (b) Triangular-lattice plane showing the disordered Nd ions. (c) Laue diffraction pattern with the X-ray beam approximately parallel to the \textit{c} axis. The inset shows a photo of the as-grown single crystal.}
	\label{xrd}
\end{figure}

\begin{figure*}
	\centering
	\includegraphics[width=1.9\columnwidth]{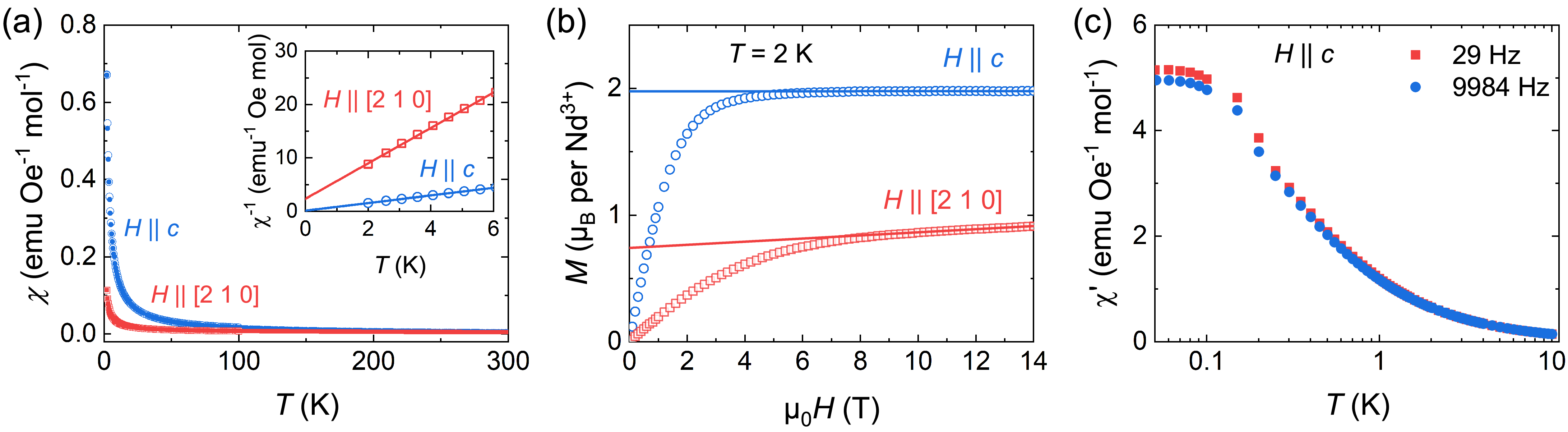}
	\caption{Magnetic properties. (a) Temperature dependence of the DC magnetic susceptibility, $\chi = M/H$, with magnetic field applied along different directions. The inset shows low-temperature Curie-Weiss fits. (b) Isothermal magnetization curves for the fields applied along the \textit{c} and [2 1 0] directions at 2 K. (c) Temperature dependence of the real component of the ac susceptibility, $\chi'$, measured at various frequencies.}
	\label{sus}
\end{figure*}

\nmao\ crystallizes into a hexagonal structure with space group $P6_3/mmc$. As shown in Fig. \ref{xrd}(a-b), the structure consists of triangular Nd magnetic layers stacked along the [001] direction and separated by nonmagnetic spacer ions. Notably, the distance between the magnetic layers amounts to approximately 10.94 \AA. Such a large separation strongly suppresses interlayer magnetic interactions, and minimizes any disorder effect arising from the nonmagnetic ions in between. The quasi-two-dimensional character can also facilitate strong quantum fluctuations.

The sharp spots of the Laue pattern indicate a high quality of the single crystals; see Fig. \ref{xrd}(c). The quality of the crystals was further characterized by (powder) single-crystal XRD measurements on the (pulverized) single crystals. No trace of any impurity phase can be detected in the powder diffraction pattern; see Fig. S2 in the SI.  Single-crystal structural refinement further reveals that $\sim$8\% of the Nd ions are displaced from the ideal $2d$ site (Nd1) toward the $6h$ site (Nd2); see Tab. S1 in the SI. This local structural disorder is consistent with previous reports on related compounds \cite{cao2024synthesis,cao2025u,kahn1981preparation,Li2024,Ma2024}.

\subsection{Magnetic Properties}

Figure \ref{sus}(a) shows the temperature dependence of the magnetic susceptibility, $\chi = M/H$, with magnetic fields applied along and perpendicular to the crystallographic \textit{c} axis. No bifurcation or difference between the zero-field-cooled (ZFC) and field-cooled (FC) curves is observed down to 2 K, indicating a paramagnetic state and the absence of spin freezing above this temperature. At low temperatures, the system exhibits a distinct magnetic anisotropy. This behavior is better visualized in the isothermal magnetization measurements as shown in Fig. \ref{sus}(b).
After a linear fit to the high-field data to subtract the van Vleck contribution, the saturated moments along the \textit{c} and [2 1 0] directions amount to 1.977(2) and 0.740(2) $\mu_\mathrm{B}$/Nd$^{3+}$, respectively.
Our inelastic neutron scattering measurements on the sister compound NdZnAl$_{11}$O$_{19}$ show that the first CEF excited state is located at about 9.4 meV \cite{Cao20252}. Therefore, the system can be considered as an effective spin-1/2 state at low temperatures. This gives rise to effective \textit{g} factors of 3.95 and 1.48 along the \textit{c} and [2~1~0] directions, respectively. The anisotropy is largely reduced compared to that of \cmao\ \cite{cao2025u} and \pmao\ \cite{cao2024synthesis}.

To extract the magnetic interactions among the ground state Kramers doublets, we perform Curie-Weiss fits, $\chi^{-1}(T) = (T - \theta_\mathrm{CW})/C$, to the inverse susceptibility in the low-temperature (2 $-$ 6 K) regime; see the inset of Fig. \ref{sus}(a). The best fits yield $\theta_\mathrm{CW}^{\|/\bot}$ of -0.228(6) and -0.71(6) K, and effective moment $\mu_\mathrm{eff}^{\|/\bot} = \sqrt{8C}$ of 3.363(2) and 1.554(7) $\mu_\mathrm{B}$/Nd, along and perpendicular to the \textit{c} axis, respectively. The negative $\theta_\mathrm{CW}$ values along both directions indicate overall antiferromagnetic interactions among the spins. Moreover, combined with the isothermal magnetization data, these results show that the single-ion anisotropy favors a moment alignment along the \textit{c} axis, while the spin-spin interactions are more significant within the \textit{ab} plane. The \textit{g} factors associated with the effective spin-1/2 state could also be evaluated by $\mu_\mathrm{eff} = g_\mathrm{eff}\sqrt{S_\mathrm{eff}(S_\mathrm{eff}+1)}$ with $S_\mathrm{eff}$ = 1/2, yielding $g_\mathrm{eff}^{\|}$ = 3.88 and $g_\mathrm{eff}^{\bot}$ = 1.79, in good agreement with that estimated from the saturated moments.
The strength of the exchange interaction $J_{\text{ex}}$ between neighboring spins can be estimated within the mean-field approximation according to $J_\mathrm{ex}^{\|/\bot} =3\theta_\text{CW}^{\|/\bot}/zS_{\text{eff}}(S_{\text{eff}} +1)$,
where $z$ = 6 is the number of nearest-neighbor Nd$^{3+}$ ions \cite{greedan2001geometrically,hamilton2014enhancement}.  Accordingly, we obtain $J_\mathrm{ex}^{\|}$ = -0.15 K and $J_\mathrm{ex}^{\bot}$ = -0.47 K.
The relatively small $J_\mathrm{ex}^{\|}$ together with the large $g_\mathrm{eff}^\|$ suggest that the ground state could be efficiently tuned by relatively weak magnetic fields applied along the \textit{c} axis, making \nmao\ a promising candidate for low-field magnetocaloric applications.

\begin{figure*}
	\centering
	\includegraphics[width=1.9\columnwidth]{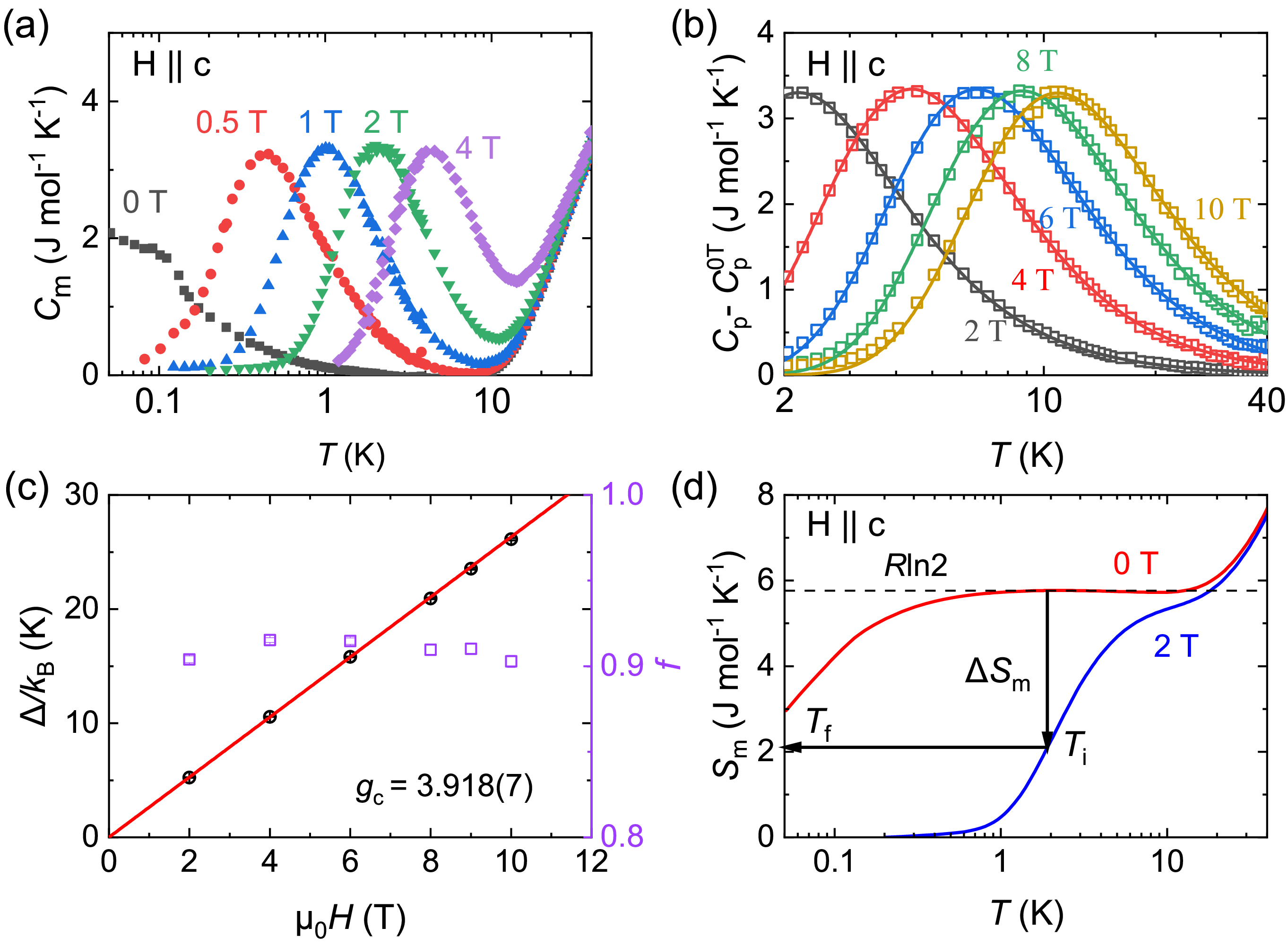}
	\caption{Thermodynamic properties. (a) Temperature dependence of the magnetic specific heat of \nmao\ measured at various magnetic fields applied along the \textit{c}-axis. Phonon contributions obtained from the nonmagnetic LaMgAl$_{11}$O$_{19}$ have been subtracted. (b) Temperature dependence of the magnetic specific heat arising from the low-lying ground state doublet of \nmao\ together with  two-level Schottky fits. (c) Evolution of the gap (left axis) and the fraction of free ions (right axis) as a function of the applied magnetic fields. The red curve represents a linear fit of $\Delta = g_c\mu_\mathrm{B}H$ to the gap, yielding $g_c$ of 3.918(7). (d) Temperature dependence of the magnetic entropy measured at different fields. The zero-field curve has been shifted vertically such that the plateau coincides with the ideal \textit{R}ln2 value.}
	\label{cp}
\end{figure*}

To clarify whether long-range order or spin freezing develops at lower temperatures, we further performed ac susceptibility measurements down to 50 mK. As shown in Fig. \ref{sus}(c), the real component of the ac susceptibility, $\chi'$, increases with decreasing temperature and tends to level off below $\sim$0.1 K, showing no signature of magnetic ordering. Moreover, the absence of a frequency-dependent cusp rules out a spin glass state \cite{Guo2016}. The finite susceptibility at low temperatures further suggests the presence of gapless low-energy magnetic excitations and persistent spin fluctuations, in line with a quantum spin liquid state.

\subsection{Specific Heat and Magnetocaloric Effect}

Specific heat can provide invaluable information about the magnetocaloric effect. Figure \ref{cp}(a) depicts the temperature dependence of the magnetic specific heat, $C_m$, under various magnetic fields. The phonon contributions have been subtracted using the nonmagnetic analogue of LaMnAl$_{11}$O$_{19}$ as a reference sample \cite{PhysRevB.43.13137}. At zero field, $C_m$ exhibits an upturn below about 1 K. However, no $\lambda$-type anomaly associated with a long-range magnetic ordering is observed down to 0.05 K. Instead, a kink is observed at $\sim$0.1 K, below which $C_m$ continues to increase slightly. The low-temperature anomaly most likely arises from electronic spin correlations, rather than a nuclear Schottky anomaly, as it is readily suppressed by a small magnetic field of 0.5 T.
At higher fields, $C_m$ evolves into a canonical Schottky anomaly with the peak intensity remaining unchanged and the peak position shifting to higher temperatures.

\begin{figure*}
	\centering
	\includegraphics[width=1.9\columnwidth]{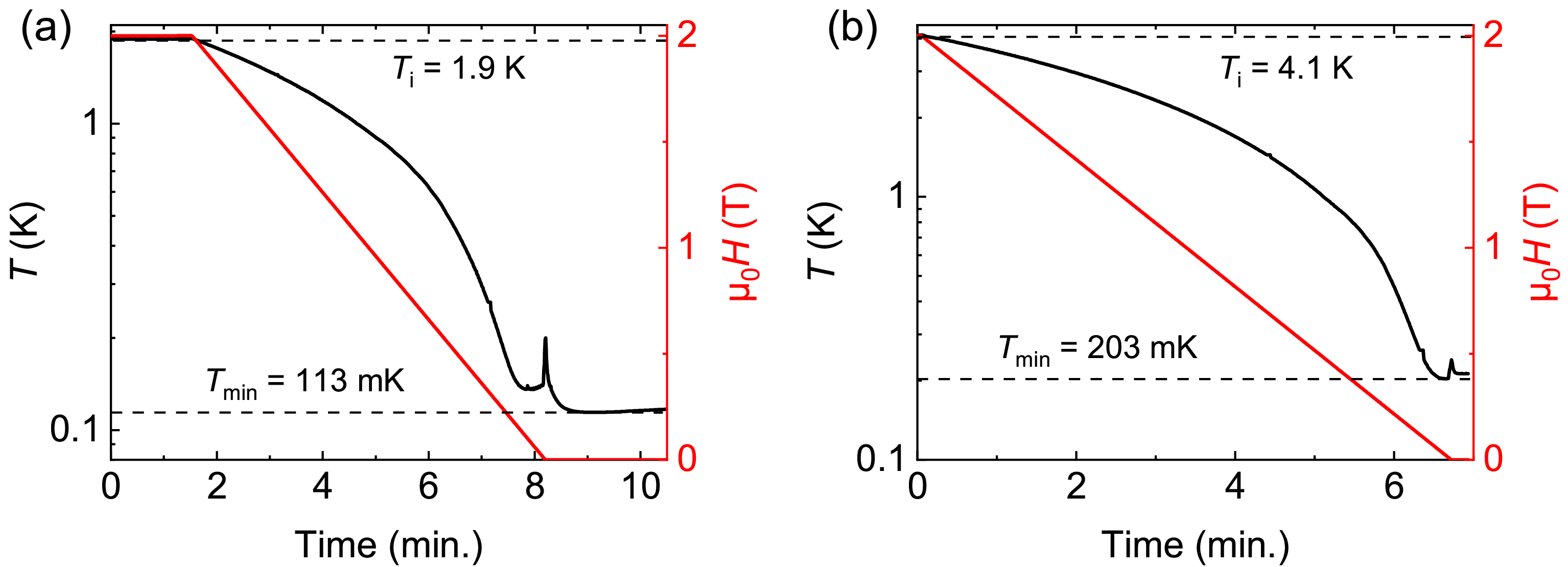}
	\caption{Quasi-adiabatic demagnetization cooling curves of \nmao, (a) starting from 2 T, 1.9 K, and (b) starting from 2 T, 4.1 K with magnetic field applied along the $c$ axis.}
	\label{mce}
\end{figure*}

Considering that the first CEF excited state is located at $\sim$9 meV, and that the interactions among the Nd$^{3+}$ spins become significant only below about 1 K, the magnetic specific heat associated with the low-lying doublet in finite magnetic fields is obtained by subtracting the zero-field data from the corresponding in-field data, as shown in Fig. \ref{cp}(b). These can be well described by a two-level Schottky model

\begin{equation}
	C_{m}=fR\left(\frac{\Delta}{k_{B}T}\right)^{2}
	\frac{\exp(\Delta/k_{B}T)}
	{\left[1+\exp(\Delta/k_{B}T)\right]^{2}},
\end{equation}
where $\Delta$ is the Zeeman gap, $R$ is the ideal gas constant, and $f$ represents the fraction of free ions. A linear fit of $\Delta = g_\mathrm{eff}\mu_{B}\mu_{0}H$ to the gap yields $g_\mathrm{eff}^{\|}=3.918(7)$, in excellent agreement with the magnetization results, confirming that the low-temperature thermodynamic response is dominated by the effective spin-$1/2$ moments.

To further evaluate the low-energy magnetic degrees of freedom, the magnetic entropy, $S_m (T)$, was obtained by integrating $C_m/T$ over temperature, as shown in Fig. \ref{cp}(d).
At zero field, only about $0.49R\ln2$ of the magnetic entropy is released from the base temperature up to 2 K, indicating that a substantial fraction of the magnetic entropy is retained at lower temperatures due to persistent spin fluctuations. In finite magnetic fields, the entropy shows a plateau at \textit{R}ln2, in line with an effective spin-1/2 ground state.
The obtained $S_m(T,H)$ phase diagram provides the basis for evaluating the magnetocaloric performance in the ADR process, as characterized by the isothermal entropy change $\Delta S_m$ under a magnetic field variation $\Delta H = H_f - H_i$ and the adiabatic temperature change $\Delta T_{\mathrm{ad}} = T_f - T_i$. Here, $\Delta S_m$ characterizes the driving force and cooling capacity of the refrigerant, while $T_f$ corresponds to the lowest attainable temperature upon reducing the magnetic field to zero. As indicated in Fig. \ref{cp}(d), starting from $T_i$ = 2 K and $H_i$ = 2 T, \nmao\ can in principle be cooled to well below 50 mK. Such a remarkable performance would surpass that of the commercial GGG \cite{PhysRevApplied.19.014038,guo2025giant}, and several other frustrated systems such as the distorted square-lattice compound NaYbGeO$_4$ ($\sim 135\ \mathrm{mK}$) \cite{PhysRevB.108.224415}, the triangular-lattice frustrated magnet KBaGd(BO$_3$)$_2$ ($\sim 122\ \mathrm{mK}$) \cite{PhysRevB.107.104402}, and the Shastry-Sutherland magnet Yb$_2$Be$_2$GeO$_7$ ($\sim 95\ \mathrm{mK}$)  \cite{PhysRevB.110.144445} according to the entropy analysis.

\begin{table*}
\caption{Comparison of key parameters of sub-Kelvin ADR materials: $T_{\mathrm{min}}$ is the attainable minimum temperature, $T_i$ is the initial temperature, $\Delta \mu_0 H$ is the change of the magnetic field, and $T_o$ is the magnetic ordering temperature.}	\label{ADRcompare}	
\centering	
\begin{tabular}{lccccc}
\hline\hline		
Material & $T_{\mathrm{min}}$ (mK) & $T_i$ (K) & $\Delta \mu_0 H$ (T) &$T_o$(K)&Lattice  \\
\hline
\GGG\ \cite{PhysRevApplied.19.014038,guo2025giant} & 800 & 1.8 & 0-2&-&hyperKagome   \\	
GdBO$_3$ \cite{lin2025quantum}&50&$\sim$0.7&0-7&1.77&triangle\\
KBaGd(BO$_3$)$_2$ \cite{PhysRevB.107.104402}& 122 & 2 & 0-5&0.263&triangle   \\	    Gd$_2$B$_2$MoO$_9$ \cite{zhang2026refrigeration}&160&2&0-9&0.79&triangle\\
NaGdP$_2$O$_7$ \cite{PhysRevB.111.064431}&220&2&0-5&0.57&quasi-one dimension\\		NH$_4$GdF$_4$\cite{guo2025giant}&$\sim$450&1.8&0-2&0.85&quasi-one dimension\\
EuCo$_2$Al$_9$ \cite{shu2026giant}&106&1.8&0-12&3.6&triangle\\
EuCl$_2$ \cite{wang2024record}&346&2&0-5&-&quasi-one dimension\\
NaYbGeO$_4$ \cite{PhysRevB.108.224415}&135&2&0-5&210&quasi-one dimension\\     	KBaYb(BO$_3$)$_2$ \cite{tokiwa2021frustrated}& 22 & 2 & 0-5&-&triangle   \\
NaYbP$_2$O$_7$ \cite{arjun2023efficient}&45&2&0-5&-&quasi-one dimension\\
YbCu$_4$Ni \cite{shimura2022magnetic}  & 170  &1.8&0-8&-&pyrochlore\\
YbPt$_2$Sn \cite{jang2015large}&220&1.75&0-4&-&triangle\\
Na$_2$BaCo(PO$_4$)$_2$ \cite{xiang2024giant}&94&2 &0-4&0.15&triangle \\
NdNi$_4$Mg \cite{zhang2025sub}& 150&2&0-9&-&FCC\\
\textbf{\nmao\ [this work]} & \textbf{113} & \textbf{1.9} & \textbf{0-2}  &-&triangle \\		
\hline\hline	
\end{tabular}
\end{table*}

Motivated by these observations, we further carried out quasi-adiabatic demagnetization measurements to quantitatively evaluate the MCE of \nmao.
As shown in Fig. \ref{mce}, with the magnetic field applied along the easy-axis direction, the sample can be cooled from an initial temperature of 4 K down to $T_\mathrm{min} \sim$ 203 mK under a magnetic field change of only 2 T. Remarkably, when the initial temperature is further reduced to 1.9 K, the system reaches a minimum temperature of about 113 mK under the same field condition. Note that a sharp temperature spike is observed when the field is reduced to around zero, which can be attributed to the flux pinning of the superconducting magnet \cite{tokiwa2021frustrated}. The continued decrease in the sample temperature after the field reaches zero is ascribed to the long spin-lattice relaxation time and low thermal conductivity of the insulating sample.
Though the actual performance is somewhat worse than that expected from the entropy analysis, the lowest attainable temperature is already far below the operating temperature for a $^3$He cryostat. To the best of our knowledge, this cooling performance surpasses that of all currently reported magnetic refrigerant materials under comparable low-field conditions, highlighting the exceptional potential of \nmao\ for low-field ADR applications.

The exceptional cooling performance of \nmao\ highlights two key ingredients for rational design of future ADR materials. The first is the retention of a large amount of entropy at low temperatures. This is best realized in rare-earth-based frustrated magnets where the interaction energy scale is usually small. Furthermore, geometric frustration or the competition between different interactions can effectively suppress any magnetic ordering down to absolute zero. This is distinct from the conventional paramagnetic salts, where magnetic order could still take place at temperatures comparable to the interaction energy scale. The second is a large effective \textit{g} factor which amplifies the Zeeman energy under a given field and therefore maximizes the field-induced entropy change. This can also be naturally fulfilled in the rare-earth systems such as Yb- and Nd-based compounds in which the spin-orbit coupling and CEF effect could result in a well-defined effective spin-1/2 state. In the Ising limit, the corresponding effective \textit{g} factor could reach values as large as 8 and 6.5 for the Yb- and Nd-based systems, respectively.

Finally, to place the refrigeration performance of \nmao\ in the context of contemporary magnetic refrigerants, we compare its key ADR parameters against those of representative paramagnetic salts and frustrated magnets in Tab. \ref{ADRcompare}.

\section{CONCLUSIONS}

In summary, we have systematically investigated the crystal structure, magnetic properties, thermodynamic behavior, and magnetocaloric performance of single-crystalline \nmao. No signature of long-range magnetic ordering is detected down to 50 mK. Specific heat measurements further reveal the retention of substantial magnetic entropy well below 50 mK, consistent with persistent fluctuations associated with a QSL state. Magnetization and susceptibility measurements show a pronounced Ising-like anisotropy with the moments preferentially aligned along the \textit{c} axis.
Quasi-adiabatic demagnetization measurements demonstrate an outstanding refrigeration performance of \nmao\ in the sub-Kelvin regime. Starting from an initial temperature of $T_i = 1.9$ K under a small field change of $\Delta\mu_0 H$ = 2 T , \nmao\ can be cooled to as low as 113 mK.

Our comprehensive study establishes \nmao\ as an exceptional candidate for low-field adiabatic demagnetization refrigeration. More broadly, these results demonstrate that preserving magnetic entropy through the suppression of magnetic ordering in frustrated materials, together with a large effective \textit{g} factor associated with the ground state doublet, constitutes a promising design principle for next-generation cryogenic magnetic refrigerants.

\section*{ACKNOWLEDGMENTS}

This work was supported by the Guangdong Basic and Applied Basic Research Foundation (Grant No. 2022B1515120020).

\bibliography{nmao} %

\begin{thebibliography}{52}%
\makeatletter
\providecommand \@ifxundefined [1]{%
 \@ifx{#1\undefined}
}%
\providecommand \@ifnum [1]{%
 \ifnum #1\expandafter \@firstoftwo
 \else \expandafter \@secondoftwo
 \fi
}%
\providecommand \@ifx [1]{%
 \ifx #1\expandafter \@firstoftwo
 \else \expandafter \@secondoftwo
 \fi
}%
\providecommand \natexlab [1]{#1}%
\providecommand \enquote  [1]{``#1''}%
\providecommand \bibnamefont  [1]{#1}%
\providecommand \bibfnamefont [1]{#1}%
\providecommand \citenamefont [1]{#1}%
\providecommand \href@noop [0]{\@secondoftwo}%
\providecommand \href [0]{\begingroup \@sanitize@url \@href}%
\providecommand \@href[1]{\@@startlink{#1}\@@href}%
\providecommand \@@href[1]{\endgroup#1\@@endlink}%
\providecommand \@sanitize@url [0]{\catcode `\\12\catcode `\$12\catcode
  `\&12\catcode `\#12\catcode `\^12\catcode `\_12\catcode `\%12\relax}%
\providecommand \@@startlink[1]{}%
\providecommand \@@endlink[0]{}%
\providecommand \url  [0]{\begingroup\@sanitize@url \@url }%
\providecommand \@url [1]{\endgroup\@href {#1}{\urlprefix }}%
\providecommand \urlprefix  [0]{URL }%
\providecommand \Eprint [0]{\href }%
\providecommand \doibase [0]{https://doi.org/}%
\providecommand \selectlanguage [0]{\@gobble}%
\providecommand \bibinfo  [0]{\@secondoftwo}%
\providecommand \bibfield  [0]{\@secondoftwo}%
\providecommand \translation [1]{[#1]}%
\providecommand \BibitemOpen [0]{}%
\providecommand \bibitemStop [0]{}%
\providecommand \bibitemNoStop [0]{.\EOS\space}%
\providecommand \EOS [0]{\spacefactor3000\relax}%
\providecommand \BibitemShut  [1]{\csname bibitem#1\endcsname}%
\let\auto@bib@innerbib\@empty
\bibitem [{\citenamefont {Gaita-Ari{\~n}o}\ \emph {et~al.}(2019)\citenamefont
  {Gaita-Ari{\~n}o}, \citenamefont {Luis}, \citenamefont {Hill},\ and\
  \citenamefont {Coronado}}]{gaita2019molecular}%
  \BibitemOpen
  \bibfield  {author} {\bibinfo {author} {\bibfnamefont {A.}~\bibnamefont
  {Gaita-Ari{\~n}o}}, \bibinfo {author} {\bibfnamefont {F.}~\bibnamefont
  {Luis}}, \bibinfo {author} {\bibfnamefont {S.}~\bibnamefont {Hill}},\ and\
  \bibinfo {author} {\bibfnamefont {E.}~\bibnamefont {Coronado}},\ }\bibfield
  {title} {\bibinfo {title} {{Molecular spins for quantum computation}},\
  }\href {https://www.nature.com/articles/s41557-019-0232-y#citeas} {\bibfield
  {journal} {\bibinfo  {journal} {Nat. Chem.}\ }\textbf {\bibinfo {volume}
  {11}},\ \bibinfo {pages} {301} (\bibinfo {year} {2019})}\BibitemShut
  {NoStop}%
\bibitem [{\citenamefont {Petit}\ \emph {et~al.}(2020)\citenamefont {Petit},
  \citenamefont {Eenink}, \citenamefont {Russ}, \citenamefont {Lawrie},
  \citenamefont {Hendrickx}, \citenamefont {Philips}, \citenamefont {Clarke},
  \citenamefont {Vandersypen},\ and\ \citenamefont
  {Veldhorst}}]{petit2020universal}%
  \BibitemOpen
  \bibfield  {author} {\bibinfo {author} {\bibfnamefont {L.}~\bibnamefont
  {Petit}}, \bibinfo {author} {\bibfnamefont {H.}~\bibnamefont {Eenink}},
  \bibinfo {author} {\bibfnamefont {M.}~\bibnamefont {Russ}}, \bibinfo {author}
  {\bibfnamefont {W.}~\bibnamefont {Lawrie}}, \bibinfo {author} {\bibfnamefont
  {N.}~\bibnamefont {Hendrickx}}, \bibinfo {author} {\bibfnamefont
  {S.}~\bibnamefont {Philips}}, \bibinfo {author} {\bibfnamefont
  {J.}~\bibnamefont {Clarke}}, \bibinfo {author} {\bibfnamefont
  {L.}~\bibnamefont {Vandersypen}},\ and\ \bibinfo {author} {\bibfnamefont
  {M.}~\bibnamefont {Veldhorst}},\ }\bibfield  {title} {\bibinfo {title}
  {{Universal quantum logic in hot silicon qubits}},\ }\href
  {https://www.nature.com/articles/s41586-020-2170-7} {\bibfield  {journal}
  {\bibinfo  {journal} {Nature}\ }\textbf {\bibinfo {volume} {580}},\ \bibinfo
  {pages} {355} (\bibinfo {year} {2020})}\BibitemShut {NoStop}%
\bibitem [{\citenamefont {Kittel}(1980)}]{kittel1980temperature}%
  \BibitemOpen
  \bibfield  {author} {\bibinfo {author} {\bibfnamefont {P.}~\bibnamefont
  {Kittel}},\ }\bibfield  {title} {\bibinfo {title} {{Temperature stabilized
  adiabatic demagnetization for space applications}},\ }\href
  {https://www.sciencedirect.com/science/article/pii/0011227580900971}
  {\bibfield  {journal} {\bibinfo  {journal} {Cryogenics}\ }\textbf {\bibinfo
  {volume} {20}},\ \bibinfo {pages} {599} (\bibinfo {year} {1980})}\BibitemShut
  {NoStop}%
\bibitem [{\citenamefont {Kittel}(1981)}]{kittel1981refrigeration}%
  \BibitemOpen
  \bibfield  {author} {\bibinfo {author} {\bibfnamefont {P.}~\bibnamefont
  {Kittel}},\ }\bibfield  {title} {\bibinfo {title} {{Refrigeration below 1 K
  in space}},\ }\href
  {https://www.sciencedirect.com/science/article/pii/0378436381908603}
  {\bibfield  {journal} {\bibinfo  {journal} {Physica B+ C}\ }\textbf {\bibinfo
  {volume} {108}},\ \bibinfo {pages} {1115} (\bibinfo {year}
  {1981})}\BibitemShut {NoStop}%
\bibitem [{\citenamefont {Zapf}\ \emph {et~al.}(2014)\citenamefont {Zapf},
  \citenamefont {Jaime},\ and\ \citenamefont {Batista}}]{RevModPhys.86.563}%
  \BibitemOpen
  \bibfield  {author} {\bibinfo {author} {\bibfnamefont {V.}~\bibnamefont
  {Zapf}}, \bibinfo {author} {\bibfnamefont {M.}~\bibnamefont {Jaime}},\ and\
  \bibinfo {author} {\bibfnamefont {C.~D.}\ \bibnamefont {Batista}},\
  }\bibfield  {title} {\bibinfo {title} {{Bose-Einstein condensation in quantum
  magnets}},\ }\href {https://doi.org/10.1103/RevModPhys.86.563} {\bibfield
  {journal} {\bibinfo  {journal} {Rev. Mod. Phys.}\ }\textbf {\bibinfo {volume}
  {86}},\ \bibinfo {pages} {563} (\bibinfo {year} {2014})}\BibitemShut
  {NoStop}%
\bibitem [{\citenamefont {Zhou}\ \emph {et~al.}(2017)\citenamefont {Zhou},
  \citenamefont {Kanoda},\ and\ \citenamefont {Ng}}]{RevModPhys.89.025003}%
  \BibitemOpen
  \bibfield  {author} {\bibinfo {author} {\bibfnamefont {Y.}~\bibnamefont
  {Zhou}}, \bibinfo {author} {\bibfnamefont {K.}~\bibnamefont {Kanoda}},\ and\
  \bibinfo {author} {\bibfnamefont {T.-K.}\ \bibnamefont {Ng}},\ }\bibfield
  {title} {\bibinfo {title} {{Quantum spin liquid states}},\ }\href
  {https://doi.org/10.1103/RevModPhys.89.025003} {\bibfield  {journal}
  {\bibinfo  {journal} {Rev. Mod. Phys.}\ }\textbf {\bibinfo {volume} {89}},\
  \bibinfo {pages} {025003} (\bibinfo {year} {2017})}\BibitemShut {NoStop}%
\bibitem [{\citenamefont {Goerbig}(2011)}]{RevModPhys.83.1193}%
  \BibitemOpen
  \bibfield  {author} {\bibinfo {author} {\bibfnamefont {M.~O.}\ \bibnamefont
  {Goerbig}},\ }\bibfield  {title} {\bibinfo {title} {{Electronic properties of
  graphene in a strong magnetic field}},\ }\href
  {https://doi.org/10.1103/RevModPhys.83.1193} {\bibfield  {journal} {\bibinfo
  {journal} {Rev. Mod. Phys.}\ }\textbf {\bibinfo {volume} {83}},\ \bibinfo
  {pages} {1193} (\bibinfo {year} {2011})}\BibitemShut {NoStop}%
\bibitem [{\citenamefont {Shirron}(2014)}]{shirron2014optimization}%
  \BibitemOpen
  \bibfield  {author} {\bibinfo {author} {\bibfnamefont {P.}~\bibnamefont
  {Shirron}},\ }\bibfield  {title} {\bibinfo {title} {{Optimization strategies
  for single-stage, multi-stage and continuous ADRs}},\ }\href
  {https://www.sciencedirect.com/science/article/pii/S0011227514000630}
  {\bibfield  {journal} {\bibinfo  {journal} {Cryogenics}\ }\textbf {\bibinfo
  {volume} {62}},\ \bibinfo {pages} {140} (\bibinfo {year} {2014})}\BibitemShut
  {NoStop}%
\bibitem [{\citenamefont {Tokiwa}\ \emph {et~al.}(2021)\citenamefont {Tokiwa},
  \citenamefont {Bachus}, \citenamefont {Kavita}, \citenamefont {Jesche},
  \citenamefont {Tsirlin},\ and\ \citenamefont
  {Gegenwart}}]{tokiwa2021frustrated}%
  \BibitemOpen
  \bibfield  {author} {\bibinfo {author} {\bibfnamefont {Y.}~\bibnamefont
  {Tokiwa}}, \bibinfo {author} {\bibfnamefont {S.}~\bibnamefont {Bachus}},
  \bibinfo {author} {\bibfnamefont {K.}~\bibnamefont {Kavita}}, \bibinfo
  {author} {\bibfnamefont {A.}~\bibnamefont {Jesche}}, \bibinfo {author}
  {\bibfnamefont {A.~A.}\ \bibnamefont {Tsirlin}},\ and\ \bibinfo {author}
  {\bibfnamefont {P.}~\bibnamefont {Gegenwart}},\ }\bibfield  {title} {\bibinfo
  {title} {{Frustrated magnet for adiabatic demagnetization cooling to
  milli-Kelvin temperatures}},\ }\href
  {https://www.nature.com/articles/s43246-021-00142-1} {\bibfield  {journal}
  {\bibinfo  {journal} {Commun. Mater.}\ }\textbf {\bibinfo {volume} {2}},\
  \bibinfo {pages} {42} (\bibinfo {year} {2021})}\BibitemShut {NoStop}%
\bibitem [{\citenamefont {Zhitomirsky}(2003)}]{PhysRevB.67.104421}%
  \BibitemOpen
  \bibfield  {author} {\bibinfo {author} {\bibfnamefont {M.~E.}\ \bibnamefont
  {Zhitomirsky}},\ }\bibfield  {title} {\bibinfo {title} {{Enhanced
  magnetocaloric effect in frustrated magnets}},\ }\href
  {https://doi.org/10.1103/PhysRevB.67.104421} {\bibfield  {journal} {\bibinfo
  {journal} {Phys. Rev. B}\ }\textbf {\bibinfo {volume} {67}},\ \bibinfo
  {pages} {104421} (\bibinfo {year} {2003})}\BibitemShut {NoStop}%
\bibitem [{\citenamefont {Sosin}\ \emph {et~al.}(2005)\citenamefont {Sosin},
  \citenamefont {Prozorova}, \citenamefont {Smirnov}, \citenamefont {Golov},
  \citenamefont {Berkutov}, \citenamefont {Petrenko}, \citenamefont
  {Balakrishnan},\ and\ \citenamefont {Zhitomirsky}}]{sosin2005magnetocaloric}%
  \BibitemOpen
  \bibfield  {author} {\bibinfo {author} {\bibfnamefont {S.}~\bibnamefont
  {Sosin}}, \bibinfo {author} {\bibfnamefont {L.}~\bibnamefont {Prozorova}},
  \bibinfo {author} {\bibfnamefont {A.}~\bibnamefont {Smirnov}}, \bibinfo
  {author} {\bibfnamefont {A.}~\bibnamefont {Golov}}, \bibinfo {author}
  {\bibfnamefont {I.}~\bibnamefont {Berkutov}}, \bibinfo {author}
  {\bibfnamefont {O.}~\bibnamefont {Petrenko}}, \bibinfo {author}
  {\bibfnamefont {G.}~\bibnamefont {Balakrishnan}},\ and\ \bibinfo {author}
  {\bibfnamefont {M.}~\bibnamefont {Zhitomirsky}},\ }\bibfield  {title}
  {\bibinfo {title} {{Magnetocaloric effect in pyrochlore antiferromagnet
  Gd$_2$Ti$_2$O$_7$}},\ }\href {https://doi.org/10.1103/PhysRevB.71.094413}
  {\bibfield  {journal} {\bibinfo  {journal} {Phys. Rev. B}\ }\textbf {\bibinfo
  {volume} {71}},\ \bibinfo {pages} {094413} (\bibinfo {year}
  {2005})}\BibitemShut {NoStop}%
\bibitem [{\citenamefont {Shu}\ \emph {et~al.}(2026)\citenamefont {Shu},
  \citenamefont {Xu}, \citenamefont {Xi}, \citenamefont {He}, \citenamefont
  {Xiang}, \citenamefont {Qu}, \citenamefont {Khalyavin}, \citenamefont
  {Manuel}, \citenamefont {Nakamura}, \citenamefont {Jiao} \emph
  {et~al.}}]{shu2026giant}%
  \BibitemOpen
  \bibfield  {author} {\bibinfo {author} {\bibfnamefont {M.}~\bibnamefont
  {Shu}}, \bibinfo {author} {\bibfnamefont {X.}~\bibnamefont {Xu}}, \bibinfo
  {author} {\bibfnamefont {N.}~\bibnamefont {Xi}}, \bibinfo {author}
  {\bibfnamefont {M.}~\bibnamefont {He}}, \bibinfo {author} {\bibfnamefont
  {J.}~\bibnamefont {Xiang}}, \bibinfo {author} {\bibfnamefont
  {G.}~\bibnamefont {Qu}}, \bibinfo {author} {\bibfnamefont {D.}~\bibnamefont
  {Khalyavin}}, \bibinfo {author} {\bibfnamefont {P.}~\bibnamefont {Manuel}},
  \bibinfo {author} {\bibfnamefont {J.~G.}\ \bibnamefont {Nakamura}}, \bibinfo
  {author} {\bibfnamefont {J.}~\bibnamefont {Jiao}}, \emph {et~al.},\
  }\bibfield  {title} {\bibinfo {title} {{Giant magnetocaloric effect and spin
  supersolid in a metallic dipolar magnet}},\ }\href
  {https://www.nature.com/articles/s41586-026-10144-z} {\bibfield  {journal}
  {\bibinfo  {journal} {Nature}\ }\textbf {\bibinfo {volume} {651}},\ \bibinfo
  {pages} {61} (\bibinfo {year} {2026})}\BibitemShut {NoStop}%
\bibitem [{\citenamefont {Treu}\ \emph {et~al.}(2024)\citenamefont {Treu},
  \citenamefont {Klinger}, \citenamefont {Oefele}, \citenamefont {Telang},
  \citenamefont {Jesche},\ and\ \citenamefont {Gegenwart}}]{treu2025utilizing}%
  \BibitemOpen
  \bibfield  {author} {\bibinfo {author} {\bibfnamefont {T.}~\bibnamefont
  {Treu}}, \bibinfo {author} {\bibfnamefont {M.}~\bibnamefont {Klinger}},
  \bibinfo {author} {\bibfnamefont {N.}~\bibnamefont {Oefele}}, \bibinfo
  {author} {\bibfnamefont {P.}~\bibnamefont {Telang}}, \bibinfo {author}
  {\bibfnamefont {A.}~\bibnamefont {Jesche}},\ and\ \bibinfo {author}
  {\bibfnamefont {P.}~\bibnamefont {Gegenwart}},\ }\bibfield  {title} {\bibinfo
  {title} {{Utilizing frustration in Gd-and Yb-based oxides for milli-Kelvin
  adiabatic demagnetization refrigeration}},\ }\href
  {https://iopscience.iop.org/article/10.1088/1361-648X/ad7dc5/meta} {\bibfield
   {journal} {\bibinfo  {journal} {J. Phys.: Condens. Matter}\ }\textbf
  {\bibinfo {volume} {37}},\ \bibinfo {pages} {013001} (\bibinfo {year}
  {2024})}\BibitemShut {NoStop}%
\bibitem [{\citenamefont {Song}\ \emph {et~al.}(2025)\citenamefont {Song},
  \citenamefont {Liu}, \citenamefont {Dong}, \citenamefont {Zhou},
  \citenamefont {Shi}, \citenamefont {Han}, \citenamefont {Ling}, \citenamefont
  {Ren}, \citenamefont {Yuan}, \citenamefont {Wang} \emph
  {et~al.}}]{song2025realization}%
  \BibitemOpen
  \bibfield  {author} {\bibinfo {author} {\bibfnamefont {F.}~\bibnamefont
  {Song}}, \bibinfo {author} {\bibfnamefont {X.}~\bibnamefont {Liu}}, \bibinfo
  {author} {\bibfnamefont {C.}~\bibnamefont {Dong}}, \bibinfo {author}
  {\bibfnamefont {J.}~\bibnamefont {Zhou}}, \bibinfo {author} {\bibfnamefont
  {X.}~\bibnamefont {Shi}}, \bibinfo {author} {\bibfnamefont {Y.}~\bibnamefont
  {Han}}, \bibinfo {author} {\bibfnamefont {L.}~\bibnamefont {Ling}}, \bibinfo
  {author} {\bibfnamefont {H.}~\bibnamefont {Ren}}, \bibinfo {author}
  {\bibfnamefont {S.}~\bibnamefont {Yuan}}, \bibinfo {author} {\bibfnamefont
  {S.}~\bibnamefont {Wang}}, \emph {et~al.},\ }\bibfield  {title} {\bibinfo
  {title} {{Realization of large magnetocaloric effect in the Kagome
  antiferromagnet Gd$_3$BWO$_9$ for Sub-Kelvin cryogenic refrigeration}},\
  }\href
  {https://iopscience.iop.org/article/10.1088/0256-307X/42/12/120706/meta}
  {\bibfield  {journal} {\bibinfo  {journal} {Chin. Phys. Lett.}\ }\textbf
  {\bibinfo {volume} {42}},\ \bibinfo {pages} {120706} (\bibinfo {year}
  {2025})}\BibitemShut {NoStop}%
\bibitem [{\citenamefont {Jesche}\ \emph {et~al.}(2023)\citenamefont {Jesche},
  \citenamefont {Winterhalter-Stocker}, \citenamefont {Hirschberger},
  \citenamefont {Bellon}, \citenamefont {Bachus}, \citenamefont {Tokiwa},
  \citenamefont {Tsirlin},\ and\ \citenamefont
  {Gegenwart}}]{PhysRevB.107.104402}%
  \BibitemOpen
  \bibfield  {author} {\bibinfo {author} {\bibfnamefont {A.}~\bibnamefont
  {Jesche}}, \bibinfo {author} {\bibfnamefont {N.}~\bibnamefont
  {Winterhalter-Stocker}}, \bibinfo {author} {\bibfnamefont {F.}~\bibnamefont
  {Hirschberger}}, \bibinfo {author} {\bibfnamefont {A.}~\bibnamefont
  {Bellon}}, \bibinfo {author} {\bibfnamefont {S.}~\bibnamefont {Bachus}},
  \bibinfo {author} {\bibfnamefont {Y.}~\bibnamefont {Tokiwa}}, \bibinfo
  {author} {\bibfnamefont {A.~A.}\ \bibnamefont {Tsirlin}},\ and\ \bibinfo
  {author} {\bibfnamefont {P.}~\bibnamefont {Gegenwart}},\ }\bibfield  {title}
  {\bibinfo {title} {{Adiabatic demagnetization cooling well below the magnetic
  ordering temperature in the triangular antiferromagnet
  ${\mathrm{KBaGd}({\mathrm{BO}}_{3})}_{2}$}},\ }\href
  {https://doi.org/10.1103/PhysRevB.107.104402} {\bibfield  {journal} {\bibinfo
   {journal} {Phys. Rev. B}\ }\textbf {\bibinfo {volume} {107}},\ \bibinfo
  {pages} {104402} (\bibinfo {year} {2023})}\BibitemShut {NoStop}%
\bibitem [{\citenamefont {Lin}\ \emph {et~al.}(2026)\citenamefont {Lin},
  \citenamefont {Zhao}, \citenamefont {Li}, \citenamefont {An}, \citenamefont
  {Guo}, \citenamefont {Wang}, \citenamefont {Pan}, \citenamefont {Wen},
  \citenamefont {Sheng}, \citenamefont {Wu} \emph {et~al.}}]{lin2025quantum}%
  \BibitemOpen
  \bibfield  {author} {\bibinfo {author} {\bibfnamefont {W.}~\bibnamefont
  {Lin}}, \bibinfo {author} {\bibfnamefont {N.}~\bibnamefont {Zhao}}, \bibinfo
  {author} {\bibfnamefont {Z.}~\bibnamefont {Li}}, \bibinfo {author}
  {\bibfnamefont {W.}~\bibnamefont {An}}, \bibinfo {author} {\bibfnamefont
  {R.}~\bibnamefont {Guo}}, \bibinfo {author} {\bibfnamefont {J.}~\bibnamefont
  {Wang}}, \bibinfo {author} {\bibfnamefont {C.}~\bibnamefont {Pan}}, \bibinfo
  {author} {\bibfnamefont {B.}~\bibnamefont {Wen}}, \bibinfo {author}
  {\bibfnamefont {J.}~\bibnamefont {Sheng}}, \bibinfo {author} {\bibfnamefont
  {L.}~\bibnamefont {Wu}}, \emph {et~al.},\ }\bibfield  {title} {\bibinfo
  {title} {{Quantum Fluctuation-enhanced Milli-Kelvin Magnetic Refrigeration in
  Triangular Lattice Magnet GdBO$_3$}},\ }\href
  {https://www.cell.com/the-innovation/fulltext/S2666-6758(26)00001-9}
  {\bibfield  {journal} {\bibinfo  {journal} {Innovation.}\ }\textbf {\bibinfo
  {volume} {7}},\ \bibinfo {pages} {101254} (\bibinfo {year}
  {2026})}\BibitemShut {NoStop}%
\bibitem [{\citenamefont {Brasiliano}\ \emph {et~al.}(2020)\citenamefont
  {Brasiliano}, \citenamefont {Duval}, \citenamefont {Marin}, \citenamefont
  {Bichaud}, \citenamefont {Brison}, \citenamefont {Zhitomirsky},\ and\
  \citenamefont {Luchier}}]{brasiliano2020ybgg}%
  \BibitemOpen
  \bibfield  {author} {\bibinfo {author} {\bibfnamefont {D.~A.~P.}\
  \bibnamefont {Brasiliano}}, \bibinfo {author} {\bibfnamefont {J.-M.}\
  \bibnamefont {Duval}}, \bibinfo {author} {\bibfnamefont {C.}~\bibnamefont
  {Marin}}, \bibinfo {author} {\bibfnamefont {E.}~\bibnamefont {Bichaud}},
  \bibinfo {author} {\bibfnamefont {J.-P.}\ \bibnamefont {Brison}}, \bibinfo
  {author} {\bibfnamefont {M.}~\bibnamefont {Zhitomirsky}},\ and\ \bibinfo
  {author} {\bibfnamefont {N.}~\bibnamefont {Luchier}},\ }\bibfield  {title}
  {\bibinfo {title} {{YbGG material for adiabatic demagnetization in the 100
  mK--3 K range}},\ }\href
  {https://www.sciencedirect.com/science/article/pii/S0011227519302413}
  {\bibfield  {journal} {\bibinfo  {journal} {Cryogenics}\ }\textbf {\bibinfo
  {volume} {105}},\ \bibinfo {pages} {103002} (\bibinfo {year}
  {2020})}\BibitemShut {NoStop}%
\bibitem [{\citenamefont {Xiang}\ \emph {et~al.}(2024)\citenamefont {Xiang},
  \citenamefont {Zhang}, \citenamefont {Gao}, \citenamefont {Schmidt},
  \citenamefont {Schmalzl}, \citenamefont {Wang}, \citenamefont {Li},
  \citenamefont {Xi}, \citenamefont {Liu}, \citenamefont {Jin} \emph
  {et~al.}}]{xiang2024giant}%
  \BibitemOpen
  \bibfield  {author} {\bibinfo {author} {\bibfnamefont {J.}~\bibnamefont
  {Xiang}}, \bibinfo {author} {\bibfnamefont {C.}~\bibnamefont {Zhang}},
  \bibinfo {author} {\bibfnamefont {Y.}~\bibnamefont {Gao}}, \bibinfo {author}
  {\bibfnamefont {W.}~\bibnamefont {Schmidt}}, \bibinfo {author} {\bibfnamefont
  {K.}~\bibnamefont {Schmalzl}}, \bibinfo {author} {\bibfnamefont {C.-W.}\
  \bibnamefont {Wang}}, \bibinfo {author} {\bibfnamefont {B.}~\bibnamefont
  {Li}}, \bibinfo {author} {\bibfnamefont {N.}~\bibnamefont {Xi}}, \bibinfo
  {author} {\bibfnamefont {X.-Y.}\ \bibnamefont {Liu}}, \bibinfo {author}
  {\bibfnamefont {H.}~\bibnamefont {Jin}}, \emph {et~al.},\ }\bibfield  {title}
  {\bibinfo {title} {{Giant magnetocaloric effect in spin supersolid candidate
  Na$_2$BaCo(PO$_4$)$_2$}},\ }\href
  {https://www.nature.com/articles/s41586-023-06885-w} {\bibfield  {journal}
  {\bibinfo  {journal} {Nature}\ }\textbf {\bibinfo {volume} {625}},\ \bibinfo
  {pages} {270} (\bibinfo {year} {2024})}\BibitemShut {NoStop}%
\bibitem [{\citenamefont {Vilches}\ and\ \citenamefont
  {Wheatley}(1966)}]{PhysRev.148.509}%
  \BibitemOpen
  \bibfield  {author} {\bibinfo {author} {\bibfnamefont {O.~E.}\ \bibnamefont
  {Vilches}}\ and\ \bibinfo {author} {\bibfnamefont {J.~C.}\ \bibnamefont
  {Wheatley}},\ }\bibfield  {title} {\bibinfo {title} {{Measurements of the
  Specific Heats of Three Magnetic Salts at Low Temperatures}},\ }\href
  {https://doi.org/10.1103/PhysRev.148.509} {\bibfield  {journal} {\bibinfo
  {journal} {Phys. Rev.}\ }\textbf {\bibinfo {volume} {148}},\ \bibinfo {pages}
  {509} (\bibinfo {year} {1966})}\BibitemShut {NoStop}%
\bibitem [{\citenamefont {Fritz}\ \emph {et~al.}(1975)\citenamefont {Fritz},
  \citenamefont {Clark},\ and\ \citenamefont {Cesarano}}]{fritz1975magnetic}%
  \BibitemOpen
  \bibfield  {author} {\bibinfo {author} {\bibfnamefont {J.}~\bibnamefont
  {Fritz}}, \bibinfo {author} {\bibfnamefont {J.}~\bibnamefont {Clark}},\ and\
  \bibinfo {author} {\bibfnamefont {J.}~\bibnamefont {Cesarano}},\ }\bibfield
  {title} {\bibinfo {title} {{Magnetic and thermodynamic properties of
  Mn(NH$_4$)$_2$(SO$_4$)$_2\cdot$6H$_2$O below 1$^\circ$ K and at fields up to
  24000 G parallel to the b axis}},\ }\href
  {https://pubs.aip.org/aip/jcp/article-abstract/62/1/200/214296} {\bibfield
  {journal} {\bibinfo  {journal} {J. Chem. Phys.}\ }\textbf {\bibinfo {volume}
  {62}},\ \bibinfo {pages} {200} (\bibinfo {year} {1975})}\BibitemShut
  {NoStop}%
\bibitem [{\citenamefont {Balents}(2010)}]{balents2010spin}%
  \BibitemOpen
  \bibfield  {author} {\bibinfo {author} {\bibfnamefont {L.}~\bibnamefont
  {Balents}},\ }\bibfield  {title} {\bibinfo {title} {{Spin liquids in
  frustrated magnets}},\ }\href {https://www.nature.com/articles/nature08917}
  {\bibfield  {journal} {\bibinfo  {journal} {Nature}\ }\textbf {\bibinfo
  {volume} {464}},\ \bibinfo {pages} {199} (\bibinfo {year}
  {2010})}\BibitemShut {NoStop}%
\bibitem [{\citenamefont {Savary}\ and\ \citenamefont
  {Balents}(2016)}]{savary2016quantum}%
  \BibitemOpen
  \bibfield  {author} {\bibinfo {author} {\bibfnamefont {L.}~\bibnamefont
  {Savary}}\ and\ \bibinfo {author} {\bibfnamefont {L.}~\bibnamefont
  {Balents}},\ }\bibfield  {title} {\bibinfo {title} {{Quantum spin liquids: a
  review}},\ }\href
  {https://iopscience.iop.org/article/10.1088/0034-4885/80/1/016502/meta}
  {\bibfield  {journal} {\bibinfo  {journal} {Rep. Prog. Phys.}\ }\textbf
  {\bibinfo {volume} {80}},\ \bibinfo {pages} {016502} (\bibinfo {year}
  {2016})}\BibitemShut {NoStop}%
\bibitem [{\citenamefont {Broholm}\ \emph {et~al.}(2020)\citenamefont
  {Broholm}, \citenamefont {Cava}, \citenamefont {Kivelson}, \citenamefont
  {Nocera}, \citenamefont {Norman},\ and\ \citenamefont
  {Senthil}}]{broholm2020quantum}%
  \BibitemOpen
  \bibfield  {author} {\bibinfo {author} {\bibfnamefont {C.}~\bibnamefont
  {Broholm}}, \bibinfo {author} {\bibfnamefont {R.~J.}\ \bibnamefont {Cava}},
  \bibinfo {author} {\bibfnamefont {S.}~\bibnamefont {Kivelson}}, \bibinfo
  {author} {\bibfnamefont {D.}~\bibnamefont {Nocera}}, \bibinfo {author}
  {\bibfnamefont {M.}~\bibnamefont {Norman}},\ and\ \bibinfo {author}
  {\bibfnamefont {T.}~\bibnamefont {Senthil}},\ }\bibfield  {title} {\bibinfo
  {title} {{Quantum spin liquids}},\ }\href
  {https://www.science.org/doi/abs/10.1126/science.aay0668} {\bibfield
  {journal} {\bibinfo  {journal} {Science}\ }\textbf {\bibinfo {volume}
  {367}},\ \bibinfo {pages} {eaay0668} (\bibinfo {year} {2020})}\BibitemShut
  {NoStop}%
\bibitem [{\citenamefont {Lee}\ \emph {et~al.}(2024)\citenamefont {Lee},
  \citenamefont {Woods}, \citenamefont {Lee}, \citenamefont {Zhang},
  \citenamefont {Choi}, \citenamefont {Scheie}, \citenamefont {Tennant},
  \citenamefont {Xing}, \citenamefont {Sefat},\ and\ \citenamefont
  {Movshovich}}]{PhysRevB.109.155129}%
  \BibitemOpen
  \bibfield  {author} {\bibinfo {author} {\bibfnamefont {S.}~\bibnamefont
  {Lee}}, \bibinfo {author} {\bibfnamefont {A.~J.}\ \bibnamefont {Woods}},
  \bibinfo {author} {\bibfnamefont {M.}~\bibnamefont {Lee}}, \bibinfo {author}
  {\bibfnamefont {S.}~\bibnamefont {Zhang}}, \bibinfo {author} {\bibfnamefont
  {E.~S.}\ \bibnamefont {Choi}}, \bibinfo {author} {\bibfnamefont {A.~O.}\
  \bibnamefont {Scheie}}, \bibinfo {author} {\bibfnamefont {D.~A.}\
  \bibnamefont {Tennant}}, \bibinfo {author} {\bibfnamefont {J.}~\bibnamefont
  {Xing}}, \bibinfo {author} {\bibfnamefont {A.~S.}\ \bibnamefont {Sefat}},\
  and\ \bibinfo {author} {\bibfnamefont {R.}~\bibnamefont {Movshovich}},\
  }\bibfield  {title} {\bibinfo {title} {{Magnetic field-temperature phase
  diagram of the spin-$\frac{1}{2}$ triangular lattice antiferromagnet
  ${\mathrm{KYbSe}}_{2}$}},\ }\href
  {https://doi.org/10.1103/PhysRevB.109.155129} {\bibfield  {journal} {\bibinfo
   {journal} {Phys. Rev. B}\ }\textbf {\bibinfo {volume} {109}},\ \bibinfo
  {pages} {155129} (\bibinfo {year} {2024})}\BibitemShut {NoStop}%
\bibitem [{\citenamefont {Liu}\ \emph {et~al.}(2022)\citenamefont {Liu},
  \citenamefont {Gao}, \citenamefont {Li}, \citenamefont {Jin}, \citenamefont
  {Xiang}, \citenamefont {Jin}, \citenamefont {Chen}, \citenamefont {Li},\ and\
  \citenamefont {Su}}]{liu2022quantum}%
  \BibitemOpen
  \bibfield  {author} {\bibinfo {author} {\bibfnamefont {X.-Y.}\ \bibnamefont
  {Liu}}, \bibinfo {author} {\bibfnamefont {Y.}~\bibnamefont {Gao}}, \bibinfo
  {author} {\bibfnamefont {H.}~\bibnamefont {Li}}, \bibinfo {author}
  {\bibfnamefont {W.}~\bibnamefont {Jin}}, \bibinfo {author} {\bibfnamefont
  {J.}~\bibnamefont {Xiang}}, \bibinfo {author} {\bibfnamefont
  {H.}~\bibnamefont {Jin}}, \bibinfo {author} {\bibfnamefont {Z.}~\bibnamefont
  {Chen}}, \bibinfo {author} {\bibfnamefont {W.}~\bibnamefont {Li}},\ and\
  \bibinfo {author} {\bibfnamefont {G.}~\bibnamefont {Su}},\ }\bibfield
  {title} {\bibinfo {title} {{Quantum spin liquid candidate as superior
  refrigerant in cascade demagnetization cooling}},\ }\href
  {https://www.nature.com/articles/s42005-022-01010-1} {\bibfield  {journal}
  {\bibinfo  {journal} {Commun. Phys.}\ }\textbf {\bibinfo {volume} {5}},\
  \bibinfo {pages} {233} (\bibinfo {year} {2022})}\BibitemShut {NoStop}%
\bibitem [{\citenamefont {Daudin}\ \emph {et~al.}(1982)\citenamefont {Daudin},
  \citenamefont {Lagnier},\ and\ \citenamefont
  {Salce}}]{daudin1982thermodynamic}%
  \BibitemOpen
  \bibfield  {author} {\bibinfo {author} {\bibfnamefont {B.}~\bibnamefont
  {Daudin}}, \bibinfo {author} {\bibfnamefont {R.}~\bibnamefont {Lagnier}},\
  and\ \bibinfo {author} {\bibfnamefont {B.}~\bibnamefont {Salce}},\ }\bibfield
   {title} {\bibinfo {title} {{Thermodynamic properties of the gadolinium
  gallium garnet, Gd$_{3}$Ga$_{5}$O$_{12}$, between 0.05 and 25 K}},\ }\href
  {https://www.sciencedirect.com/science/article/pii/0304885382900920}
  {\bibfield  {journal} {\bibinfo  {journal} {J. Magn. Magn. Mater.}\ }\textbf
  {\bibinfo {volume} {27}},\ \bibinfo {pages} {315} (\bibinfo {year}
  {1982})}\BibitemShut {NoStop}%
\bibitem [{\citenamefont {Dai}\ \emph {et~al.}(1988)\citenamefont {Dai},
  \citenamefont {Gmelin},\ and\ \citenamefont
  {Kremer}}]{dai1988magnetothermal}%
  \BibitemOpen
  \bibfield  {author} {\bibinfo {author} {\bibfnamefont {W.}~\bibnamefont
  {Dai}}, \bibinfo {author} {\bibfnamefont {E.}~\bibnamefont {Gmelin}},\ and\
  \bibinfo {author} {\bibfnamefont {R.}~\bibnamefont {Kremer}},\ }\bibfield
  {title} {\bibinfo {title} {{Magnetothermal properties of sintered
  Gd$_{3}$Ga$_{5}$O$_{12}$}},\ }\href
  {https://iopscience.iop.org/article/10.1088/0022-3727/21/4/014/meta}
  {\bibfield  {journal} {\bibinfo  {journal} {J. Phys. D: Appl. Phys.}\
  }\textbf {\bibinfo {volume} {21}},\ \bibinfo {pages} {628} (\bibinfo {year}
  {1988})}\BibitemShut {NoStop}%
\bibitem [{\citenamefont {Ashtar}\ \emph {et~al.}(2019)\citenamefont {Ashtar},
  \citenamefont {Marwat}, \citenamefont {Gao}, \citenamefont {Zhang},
  \citenamefont {Pi}, \citenamefont {Yuan},\ and\ \citenamefont
  {Tian}}]{ashtar2019}%
  \BibitemOpen
  \bibfield  {author} {\bibinfo {author} {\bibfnamefont {M.}~\bibnamefont
  {Ashtar}}, \bibinfo {author} {\bibfnamefont {M.}~\bibnamefont {Marwat}},
  \bibinfo {author} {\bibfnamefont {Y.}~\bibnamefont {Gao}}, \bibinfo {author}
  {\bibfnamefont {Z.}~\bibnamefont {Zhang}}, \bibinfo {author} {\bibfnamefont
  {L.}~\bibnamefont {Pi}}, \bibinfo {author} {\bibfnamefont {S.}~\bibnamefont
  {Yuan}},\ and\ \bibinfo {author} {\bibfnamefont {Z.}~\bibnamefont {Tian}},\
  }\bibfield  {title} {\bibinfo {title} {{REZnAl$_{11}$O$_{19}$ (RE = Pr, Nd,
  Sm--Tb): a new family of ideal 2D triangular lattice frustrated magnets}},\
  }\href {https://pubs.rsc.org/en/content/articlehtml/2019/tc/c9tc02643f}
  {\bibfield  {journal} {\bibinfo  {journal} {J. Mater. Chem. C}\ }\textbf
  {\bibinfo {volume} {7}},\ \bibinfo {pages} {10073} (\bibinfo {year}
  {2019})}\BibitemShut {NoStop}%
\bibitem [{\citenamefont {Bu}\ \emph {et~al.}(2022)\citenamefont {Bu},
  \citenamefont {Ashtar}, \citenamefont {Shiroka}, \citenamefont {Walker},
  \citenamefont {Fu}, \citenamefont {Zhao}, \citenamefont {Gardner},
  \citenamefont {Chen}, \citenamefont {Tian},\ and\ \citenamefont
  {Guo}}]{Bu2022}%
  \BibitemOpen
  \bibfield  {author} {\bibinfo {author} {\bibfnamefont {H.}~\bibnamefont
  {Bu}}, \bibinfo {author} {\bibfnamefont {M.}~\bibnamefont {Ashtar}}, \bibinfo
  {author} {\bibfnamefont {T.}~\bibnamefont {Shiroka}}, \bibinfo {author}
  {\bibfnamefont {H.~C.}\ \bibnamefont {Walker}}, \bibinfo {author}
  {\bibfnamefont {Z.}~\bibnamefont {Fu}}, \bibinfo {author} {\bibfnamefont
  {J.}~\bibnamefont {Zhao}}, \bibinfo {author} {\bibfnamefont {J.~S.}\
  \bibnamefont {Gardner}}, \bibinfo {author} {\bibfnamefont {G.}~\bibnamefont
  {Chen}}, \bibinfo {author} {\bibfnamefont {Z.}~\bibnamefont {Tian}},\ and\
  \bibinfo {author} {\bibfnamefont {H.}~\bibnamefont {Guo}},\ }\bibfield
  {title} {\bibinfo {title} {{Gapless triangular-lattice spin-liquid candidate
  ${\mathrm{PrZnAl}}_{11}{\mathrm{O}}_{19}$}},\ }\href
  {https://doi.org/10.1103/PhysRevB.106.134428} {\bibfield  {journal} {\bibinfo
   {journal} {Phys. Rev. B}\ }\textbf {\bibinfo {volume} {106}},\ \bibinfo
  {pages} {134428} (\bibinfo {year} {2022})}\BibitemShut {NoStop}%
\bibitem [{\citenamefont {Cao}\ \emph {et~al.}(2024)\citenamefont {Cao},
  \citenamefont {Bu}, \citenamefont {Fu}, \citenamefont {Zhao}, \citenamefont
  {Gardner}, \citenamefont {Ouyang}, \citenamefont {Tian}, \citenamefont {Li},\
  and\ \citenamefont {Guo}}]{cao2024synthesis}%
  \BibitemOpen
  \bibfield  {author} {\bibinfo {author} {\bibfnamefont {Y.}~\bibnamefont
  {Cao}}, \bibinfo {author} {\bibfnamefont {H.}~\bibnamefont {Bu}}, \bibinfo
  {author} {\bibfnamefont {Z.}~\bibnamefont {Fu}}, \bibinfo {author}
  {\bibfnamefont {J.}~\bibnamefont {Zhao}}, \bibinfo {author} {\bibfnamefont
  {J.~S.}\ \bibnamefont {Gardner}}, \bibinfo {author} {\bibfnamefont
  {Z.}~\bibnamefont {Ouyang}}, \bibinfo {author} {\bibfnamefont
  {Z.}~\bibnamefont {Tian}}, \bibinfo {author} {\bibfnamefont {Z.}~\bibnamefont
  {Li}},\ and\ \bibinfo {author} {\bibfnamefont {H.}~\bibnamefont {Guo}},\
  }\bibfield  {title} {\bibinfo {title} {{Synthesis, disorder and Ising
  anisotropy in a new spin liquid candidate PrMgAl$_{11}$O$_{19}$}},\ }\href
  {https://iopscience.iop.org/article/10.1088/2752-5724/ad4a93/meta} {\bibfield
   {journal} {\bibinfo  {journal} {Mater. Futures}\ }\textbf {\bibinfo {volume}
  {3}},\ \bibinfo {pages} {035201} (\bibinfo {year} {2024})}\BibitemShut
  {NoStop}%
\bibitem [{\citenamefont {Cao}\ \emph {et~al.}(2025{\natexlab{a}})\citenamefont
  {Cao}, \citenamefont {Koda}, \citenamefont {Le}, \citenamefont {Pomjakushin},
  \citenamefont {Liu}, \citenamefont {Fu}, \citenamefont {Li}, \citenamefont
  {Zhao}, \citenamefont {Tian},\ and\ \citenamefont {Guo}}]{cao2025u}%
  \BibitemOpen
  \bibfield  {author} {\bibinfo {author} {\bibfnamefont {Y.}~\bibnamefont
  {Cao}}, \bibinfo {author} {\bibfnamefont {A.}~\bibnamefont {Koda}}, \bibinfo
  {author} {\bibfnamefont {M.}~\bibnamefont {Le}}, \bibinfo {author}
  {\bibfnamefont {V.}~\bibnamefont {Pomjakushin}}, \bibinfo {author}
  {\bibfnamefont {B.}~\bibnamefont {Liu}}, \bibinfo {author} {\bibfnamefont
  {Z.}~\bibnamefont {Fu}}, \bibinfo {author} {\bibfnamefont {Z.}~\bibnamefont
  {Li}}, \bibinfo {author} {\bibfnamefont {J.}~\bibnamefont {Zhao}}, \bibinfo
  {author} {\bibfnamefont {Z.}~\bibnamefont {Tian}},\ and\ \bibinfo {author}
  {\bibfnamefont {H.}~\bibnamefont {Guo}},\ }\bibfield  {title} {\bibinfo
  {title} {{U(1) Dirac quantum spin liquid candidate in triangular-lattice
  antiferromagnet CeMgAl$_{11}$O$_{19}$}},\ }\href
  {https://link.springer.com/article/10.1007/s11433-024-2634-9} {\bibfield
  {journal} {\bibinfo  {journal} {Sci. China: Phys. Mech. Astron.}\ }\textbf
  {\bibinfo {volume} {68}},\ \bibinfo {pages} {1} (\bibinfo {year}
  {2025}{\natexlab{a}})}\BibitemShut {NoStop}%
\bibitem [{\citenamefont {Li}\ \emph {et~al.}(2024)\citenamefont {Li},
  \citenamefont {Rutherford}, \citenamefont {Wang}, \citenamefont {Liang},
  \citenamefont {Li}, \citenamefont {Zhang}, \citenamefont {Wang},
  \citenamefont {Xie}, \citenamefont {Zhou},\ and\ \citenamefont
  {Sun}}]{Li2024}%
  \BibitemOpen
  \bibfield  {author} {\bibinfo {author} {\bibfnamefont {N.}~\bibnamefont
  {Li}}, \bibinfo {author} {\bibfnamefont {A.}~\bibnamefont {Rutherford}},
  \bibinfo {author} {\bibfnamefont {Y.~Y.}\ \bibnamefont {Wang}}, \bibinfo
  {author} {\bibfnamefont {H.}~\bibnamefont {Liang}}, \bibinfo {author}
  {\bibfnamefont {Q.~J.}\ \bibnamefont {Li}}, \bibinfo {author} {\bibfnamefont
  {Z.~J.}\ \bibnamefont {Zhang}}, \bibinfo {author} {\bibfnamefont
  {H.}~\bibnamefont {Wang}}, \bibinfo {author} {\bibfnamefont {W.}~\bibnamefont
  {Xie}}, \bibinfo {author} {\bibfnamefont {H.~D.}\ \bibnamefont {Zhou}},\ and\
  \bibinfo {author} {\bibfnamefont {X.~F.}\ \bibnamefont {Sun}},\ }\bibfield
  {title} {\bibinfo {title} {{Ising-type quantum spin liquid state in
  ${\mathrm{PrMgAl}}_{11}{\mathrm{O}}_{19}$}},\ }\href
  {https://doi.org/10.1103/PhysRevB.110.134401} {\bibfield  {journal} {\bibinfo
   {journal} {Phys. Rev. B}\ }\textbf {\bibinfo {volume} {110}},\ \bibinfo
  {pages} {134401} (\bibinfo {year} {2024})}\BibitemShut {NoStop}%
\bibitem [{\citenamefont {Ma}\ \emph {et~al.}(2024)\citenamefont {Ma},
  \citenamefont {Zheng}, \citenamefont {Chen}, \citenamefont {Xu},
  \citenamefont {Dong}, \citenamefont {Wang}, \citenamefont {Du}, \citenamefont
  {Embs}, \citenamefont {Li}, \citenamefont {Li}, \citenamefont {Zhang},
  \citenamefont {Liu}, \citenamefont {Zhong}, \citenamefont {Liu},\ and\
  \citenamefont {Wen}}]{Ma2024}%
  \BibitemOpen
  \bibfield  {author} {\bibinfo {author} {\bibfnamefont {Z.}~\bibnamefont
  {Ma}}, \bibinfo {author} {\bibfnamefont {S.}~\bibnamefont {Zheng}}, \bibinfo
  {author} {\bibfnamefont {Y.}~\bibnamefont {Chen}}, \bibinfo {author}
  {\bibfnamefont {R.}~\bibnamefont {Xu}}, \bibinfo {author} {\bibfnamefont
  {Z.-Y.}\ \bibnamefont {Dong}}, \bibinfo {author} {\bibfnamefont
  {J.}~\bibnamefont {Wang}}, \bibinfo {author} {\bibfnamefont {H.}~\bibnamefont
  {Du}}, \bibinfo {author} {\bibfnamefont {J.~P.}\ \bibnamefont {Embs}},
  \bibinfo {author} {\bibfnamefont {S.}~\bibnamefont {Li}}, \bibinfo {author}
  {\bibfnamefont {Y.}~\bibnamefont {Li}}, \bibinfo {author} {\bibfnamefont
  {Y.}~\bibnamefont {Zhang}}, \bibinfo {author} {\bibfnamefont
  {M.}~\bibnamefont {Liu}}, \bibinfo {author} {\bibfnamefont {R.}~\bibnamefont
  {Zhong}}, \bibinfo {author} {\bibfnamefont {J.-M.}\ \bibnamefont {Liu}},\
  and\ \bibinfo {author} {\bibfnamefont {J.}~\bibnamefont {Wen}},\ }\bibfield
  {title} {\bibinfo {title} {{Possible gapless quantum spin liquid behavior in
  the triangular-lattice Ising antiferromagnet
  ${\mathrm{PrMgAl}}_{11}{\mathrm{O}}_{19}$}},\ }\href
  {https://doi.org/10.1103/PhysRevB.109.165143} {\bibfield  {journal} {\bibinfo
   {journal} {Phys. Rev. B}\ }\textbf {\bibinfo {volume} {109}},\ \bibinfo
  {pages} {165143} (\bibinfo {year} {2024})}\BibitemShut {NoStop}%
\bibitem [{\citenamefont {Pet{\v{r}}{\'\i}{\v{c}}ek}\ \emph
  {et~al.}(2023)\citenamefont {Pet{\v{r}}{\'\i}{\v{c}}ek}, \citenamefont
  {Palatinus}, \citenamefont {Pl{\'a}{\v{s}}il},\ and\ \citenamefont
  {Du{\v{s}}ek}}]{petvrivcek2023jana2020}%
  \BibitemOpen
  \bibfield  {author} {\bibinfo {author} {\bibfnamefont {V.}~\bibnamefont
  {Pet{\v{r}}{\'\i}{\v{c}}ek}}, \bibinfo {author} {\bibfnamefont
  {L.}~\bibnamefont {Palatinus}}, \bibinfo {author} {\bibfnamefont
  {J.}~\bibnamefont {Pl{\'a}{\v{s}}il}},\ and\ \bibinfo {author} {\bibfnamefont
  {M.}~\bibnamefont {Du{\v{s}}ek}},\ }\bibfield  {title} {\bibinfo {title}
  {{Jana2020--a new version of the crystallographic computing system Jana}},\
  }\href
  {https://www.degruyterbrill.com/document/doi/10.1515/zkri-2023-0005/html}
  {\bibfield  {journal} {\bibinfo  {journal} {Z. Kristallogr. Cryst. Mater.}\
  }\textbf {\bibinfo {volume} {238}},\ \bibinfo {pages} {271} (\bibinfo {year}
  {2023})}\BibitemShut {NoStop}%
\bibitem [{\citenamefont
  {Rodr{\'\i}guez-Carvajal}(1993)}]{rodriguez1993recent}%
  \BibitemOpen
  \bibfield  {author} {\bibinfo {author} {\bibfnamefont {J.}~\bibnamefont
  {Rodr{\'\i}guez-Carvajal}},\ }\bibfield  {title} {\bibinfo {title} {{Recent
  advances in magnetic structure determination by neutron powder
  diffraction}},\ }\href
  {https://www.sciencedirect.com/science/article/pii/092145269390108I}
  {\bibfield  {journal} {\bibinfo  {journal} {Physica B}\ }\textbf {\bibinfo
  {volume} {192}},\ \bibinfo {pages} {55} (\bibinfo {year} {1993})}\BibitemShut
  {NoStop}%
\bibitem [{\citenamefont {Kahn}\ \emph {et~al.}(1981)\citenamefont {Kahn},
  \citenamefont {Lejus}, \citenamefont {Madsac}, \citenamefont {Thery},
  \citenamefont {Vivien},\ and\ \citenamefont {Bernier}}]{kahn1981preparation}%
  \BibitemOpen
  \bibfield  {author} {\bibinfo {author} {\bibfnamefont {A.}~\bibnamefont
  {Kahn}}, \bibinfo {author} {\bibfnamefont {A.-M.}\ \bibnamefont {Lejus}},
  \bibinfo {author} {\bibfnamefont {M.}~\bibnamefont {Madsac}}, \bibinfo
  {author} {\bibfnamefont {J.}~\bibnamefont {Thery}}, \bibinfo {author}
  {\bibfnamefont {D.}~\bibnamefont {Vivien}},\ and\ \bibinfo {author}
  {\bibfnamefont {J.}~\bibnamefont {Bernier}},\ }\bibfield  {title} {\bibinfo
  {title} {{Preparation, structure, optical, and magnetic properties of
  lanthanide aluminate single crystals (LnMAl$_{11}$O$_{19}$)}},\ }\href
  {https://pubs.aip.org/aip/jap/article-abstract/52/11/6864/10188} {\bibfield
  {journal} {\bibinfo  {journal} {J. Appl. Phys.}\ }\textbf {\bibinfo {volume}
  {52}},\ \bibinfo {pages} {6864} (\bibinfo {year} {1981})}\BibitemShut
  {NoStop}%
\bibitem [{\citenamefont {Cao}\ \emph {et~al.}(2025{\natexlab{b}})\citenamefont
  {Cao}, \citenamefont {Bu}, \citenamefont {Shiroka}, \citenamefont {Walker},
  \citenamefont {Fu}, \citenamefont {Tian}, \citenamefont {Zhao},\ and\
  \citenamefont {Guo}}]{Cao20252}%
  \BibitemOpen
  \bibfield  {author} {\bibinfo {author} {\bibfnamefont {Y.}~\bibnamefont
  {Cao}}, \bibinfo {author} {\bibfnamefont {H.}~\bibnamefont {Bu}}, \bibinfo
  {author} {\bibfnamefont {T.}~\bibnamefont {Shiroka}}, \bibinfo {author}
  {\bibfnamefont {H.~C.}\ \bibnamefont {Walker}}, \bibinfo {author}
  {\bibfnamefont {Z.}~\bibnamefont {Fu}}, \bibinfo {author} {\bibfnamefont
  {Z.}~\bibnamefont {Tian}}, \bibinfo {author} {\bibfnamefont {J.}~\bibnamefont
  {Zhao}},\ and\ \bibinfo {author} {\bibfnamefont {H.}~\bibnamefont {Guo}},\
  }\bibfield  {title} {\bibinfo {title} {{Magnetic ground state and persistent
  spin fluctuations in the triangular-lattice antiferromagnet
  ${\mathrm{NdZnAl}}_{11}{\mathrm{O}}_{19}$}},\ }\href
  {https://doi.org/10.1103/tnwb-9hv8} {\bibfield  {journal} {\bibinfo
  {journal} {Phys. Rev. B}\ }\textbf {\bibinfo {volume} {112}},\ \bibinfo
  {pages} {144409} (\bibinfo {year} {2025}{\natexlab{b}})}\BibitemShut
  {NoStop}%
\bibitem [{\citenamefont {Greedan}(2001)}]{greedan2001geometrically}%
  \BibitemOpen
  \bibfield  {author} {\bibinfo {author} {\bibfnamefont {J.~E.}\ \bibnamefont
  {Greedan}},\ }\bibfield  {title} {\bibinfo {title} {{Geometrically frustrated
  magnetic materials}},\ }\href
  {https://pubs.rsc.org/en/content/articlehtml/2001/jm/b003682j} {\bibfield
  {journal} {\bibinfo  {journal} {J. Mater. Chem.}\ }\textbf {\bibinfo {volume}
  {11}},\ \bibinfo {pages} {37} (\bibinfo {year} {2001})}\BibitemShut {NoStop}%
\bibitem [{\citenamefont {Hamilton}\ \emph {et~al.}(2014)\citenamefont
  {Hamilton}, \citenamefont {Lampronti}, \citenamefont {Rowley},\ and\
  \citenamefont {Dutton}}]{hamilton2014enhancement}%
  \BibitemOpen
  \bibfield  {author} {\bibinfo {author} {\bibfnamefont {A.~S.}\ \bibnamefont
  {Hamilton}}, \bibinfo {author} {\bibfnamefont {G.}~\bibnamefont {Lampronti}},
  \bibinfo {author} {\bibfnamefont {S.}~\bibnamefont {Rowley}},\ and\ \bibinfo
  {author} {\bibfnamefont {S.}~\bibnamefont {Dutton}},\ }\bibfield  {title}
  {\bibinfo {title} {{Enhancement of the magnetocaloric effect driven by
  changes in the crystal structure of Al-doped GGG,
  Gd$_3$Ga$_{5-x}$Al$_x$O$_{12}$ (0 $\leq x \leq$ 5)}},\ }\href
  {https://iopscience.iop.org/article/10.1088/0953-8984/26/11/116001/meta}
  {\bibfield  {journal} {\bibinfo  {journal} {J. Phys. Condens. Matter}\
  }\textbf {\bibinfo {volume} {26}},\ \bibinfo {pages} {116001} (\bibinfo
  {year} {2014})}\BibitemShut {NoStop}%
\bibitem [{\citenamefont {Guo}\ \emph {et~al.}(2016)\citenamefont {Guo},
  \citenamefont {Manna}, \citenamefont {Luetkens}, \citenamefont {Hoelzel},\
  and\ \citenamefont {Komarek}}]{Guo2016}%
  \BibitemOpen
  \bibfield  {author} {\bibinfo {author} {\bibfnamefont {H.}~\bibnamefont
  {Guo}}, \bibinfo {author} {\bibfnamefont {K.}~\bibnamefont {Manna}}, \bibinfo
  {author} {\bibfnamefont {H.}~\bibnamefont {Luetkens}}, \bibinfo {author}
  {\bibfnamefont {M.}~\bibnamefont {Hoelzel}},\ and\ \bibinfo {author}
  {\bibfnamefont {A.~C.}\ \bibnamefont {Komarek}},\ }\bibfield  {title}
  {\bibinfo {title} {{Spin glass behavior in
  ${\mathrm{LaCo}}_{1\ensuremath{-}x}{\mathrm{Rh}}_{x}{\mathrm{O}}_{3}$
  ($x=0.4$, 0.5, and 0.6)}},\ }\href
  {https://doi.org/10.1103/PhysRevB.94.205128} {\bibfield  {journal} {\bibinfo
  {journal} {Phys. Rev. B}\ }\textbf {\bibinfo {volume} {94}},\ \bibinfo
  {pages} {205128} (\bibinfo {year} {2016})}\BibitemShut {NoStop}%
\bibitem [{\citenamefont {Bouvier}\ \emph {et~al.}(1991)\citenamefont
  {Bouvier}, \citenamefont {Lethuillier},\ and\ \citenamefont
  {Schmitt}}]{PhysRevB.43.13137}%
  \BibitemOpen
  \bibfield  {author} {\bibinfo {author} {\bibfnamefont {M.}~\bibnamefont
  {Bouvier}}, \bibinfo {author} {\bibfnamefont {P.}~\bibnamefont
  {Lethuillier}},\ and\ \bibinfo {author} {\bibfnamefont {D.}~\bibnamefont
  {Schmitt}},\ }\bibfield  {title} {\bibinfo {title} {{Specific heat in some
  gadolinium compounds. I. Experimental}},\ }\href
  {https://doi.org/10.1103/PhysRevB.43.13137} {\bibfield  {journal} {\bibinfo
  {journal} {Phys. Rev. B}\ }\textbf {\bibinfo {volume} {43}},\ \bibinfo
  {pages} {13137} (\bibinfo {year} {1991})}\BibitemShut {NoStop}%
\bibitem [{\citenamefont {Kleinhans}\ \emph {et~al.}(2023)\citenamefont
  {Kleinhans}, \citenamefont {Eibensteiner}, \citenamefont {Leiner},
  \citenamefont {Resch}, \citenamefont {Worch}, \citenamefont {Wilde},
  \citenamefont {Spallek}, \citenamefont {Regnat},\ and\ \citenamefont
  {Pfleiderer}}]{PhysRevApplied.19.014038}%
  \BibitemOpen
  \bibfield  {author} {\bibinfo {author} {\bibfnamefont {M.}~\bibnamefont
  {Kleinhans}}, \bibinfo {author} {\bibfnamefont {K.}~\bibnamefont
  {Eibensteiner}}, \bibinfo {author} {\bibfnamefont {J.}~\bibnamefont
  {Leiner}}, \bibinfo {author} {\bibfnamefont {C.}~\bibnamefont {Resch}},
  \bibinfo {author} {\bibfnamefont {L.}~\bibnamefont {Worch}}, \bibinfo
  {author} {\bibfnamefont {M.}~\bibnamefont {Wilde}}, \bibinfo {author}
  {\bibfnamefont {J.}~\bibnamefont {Spallek}}, \bibinfo {author} {\bibfnamefont
  {A.}~\bibnamefont {Regnat}},\ and\ \bibinfo {author} {\bibfnamefont
  {C.}~\bibnamefont {Pfleiderer}},\ }\bibfield  {title} {\bibinfo {title}
  {{Magnetocaloric Properties of ${R}_{3}{\mathrm{Ga}}_{5}{\mathrm{O}}_{12}$
  ($R=\text{Tb, Gd, Nd, Dy}$)}},\ }\href
  {https://doi.org/10.1103/PhysRevApplied.19.014038} {\bibfield  {journal}
  {\bibinfo  {journal} {Phys. Rev. Appl.}\ }\textbf {\bibinfo {volume} {19}},\
  \bibinfo {pages} {014038} (\bibinfo {year} {2023})}\BibitemShut {NoStop}%
\bibitem [{\citenamefont {Guo}\ \emph {et~al.}(2025)\citenamefont {Guo},
  \citenamefont {Ren}, \citenamefont {Liu}, \citenamefont {Yao}, \citenamefont
  {Xiang}, \citenamefont {Zhang}, \citenamefont {Wang}, \citenamefont {Kumara},
  \citenamefont {Wang}, \citenamefont {Li} \emph {et~al.}}]{guo2025giant}%
  \BibitemOpen
  \bibfield  {author} {\bibinfo {author} {\bibfnamefont {Q.}~\bibnamefont
  {Guo}}, \bibinfo {author} {\bibfnamefont {W.}~\bibnamefont {Ren}}, \bibinfo
  {author} {\bibfnamefont {P.}~\bibnamefont {Liu}}, \bibinfo {author}
  {\bibfnamefont {J.}~\bibnamefont {Yao}}, \bibinfo {author} {\bibfnamefont
  {J.}~\bibnamefont {Xiang}}, \bibinfo {author} {\bibfnamefont
  {K.}~\bibnamefont {Zhang}}, \bibinfo {author} {\bibfnamefont
  {Y.}~\bibnamefont {Wang}}, \bibinfo {author} {\bibfnamefont {L.~S.~R.}\
  \bibnamefont {Kumara}}, \bibinfo {author} {\bibfnamefont {X.}~\bibnamefont
  {Wang}}, \bibinfo {author} {\bibfnamefont {W.}~\bibnamefont {Li}}, \emph
  {et~al.},\ }\bibfield  {title} {\bibinfo {title} {{Giant Low-Field
  Magnetocaloric Effect at Sub-Kelvin Temperatures in Ferromagnetic
  NH$_4$GdF$_4$}},\ }\href {https://pubs.acs.org/doi/abs/10.1021/jacs.5c10979}
  {\bibfield  {journal} {\bibinfo  {journal} {J. Am. Chem. Soc.}\ }\textbf
  {\bibinfo {volume} {147}},\ \bibinfo {pages} {34862} (\bibinfo {year}
  {2025})}\BibitemShut {NoStop}%
\bibitem [{\citenamefont {Arjun}\ \emph
  {et~al.}(2023{\natexlab{a}})\citenamefont {Arjun}, \citenamefont {Ranjith},
  \citenamefont {Jesche}, \citenamefont {Hirschberger}, \citenamefont {Sarma},\
  and\ \citenamefont {Gegenwart}}]{PhysRevB.108.224415}%
  \BibitemOpen
  \bibfield  {author} {\bibinfo {author} {\bibfnamefont {U.}~\bibnamefont
  {Arjun}}, \bibinfo {author} {\bibfnamefont {K.~M.}\ \bibnamefont {Ranjith}},
  \bibinfo {author} {\bibfnamefont {A.}~\bibnamefont {Jesche}}, \bibinfo
  {author} {\bibfnamefont {F.}~\bibnamefont {Hirschberger}}, \bibinfo {author}
  {\bibfnamefont {D.~D.}\ \bibnamefont {Sarma}},\ and\ \bibinfo {author}
  {\bibfnamefont {P.}~\bibnamefont {Gegenwart}},\ }\bibfield  {title} {\bibinfo
  {title} {{Adiabatic demagnetization refrigeration to millikelvin temperatures
  with the distorted square lattice magnet ${\mathrm{NaYbGeO}}_{4}$}},\ }\href
  {https://doi.org/10.1103/PhysRevB.108.224415} {\bibfield  {journal} {\bibinfo
   {journal} {Phys. Rev. B}\ }\textbf {\bibinfo {volume} {108}},\ \bibinfo
  {pages} {224415} (\bibinfo {year} {2023}{\natexlab{a}})}\BibitemShut
  {NoStop}%
\bibitem [{\citenamefont {Liu}\ \emph {et~al.}(2024)\citenamefont {Liu},
  \citenamefont {Zhou}, \citenamefont {Wang}, \citenamefont {Cao},
  \citenamefont {Song}, \citenamefont {Han}, \citenamefont {Li}, \citenamefont
  {Tong}, \citenamefont {Xia}, \citenamefont {Ouyang}, \citenamefont {Zhao},
  \citenamefont {Guo},\ and\ \citenamefont {Tian}}]{PhysRevB.110.144445}%
  \BibitemOpen
  \bibfield  {author} {\bibinfo {author} {\bibfnamefont {A.}~\bibnamefont
  {Liu}}, \bibinfo {author} {\bibfnamefont {J.}~\bibnamefont {Zhou}}, \bibinfo
  {author} {\bibfnamefont {L.}~\bibnamefont {Wang}}, \bibinfo {author}
  {\bibfnamefont {Y.}~\bibnamefont {Cao}}, \bibinfo {author} {\bibfnamefont
  {F.}~\bibnamefont {Song}}, \bibinfo {author} {\bibfnamefont {Y.}~\bibnamefont
  {Han}}, \bibinfo {author} {\bibfnamefont {J.}~\bibnamefont {Li}}, \bibinfo
  {author} {\bibfnamefont {W.}~\bibnamefont {Tong}}, \bibinfo {author}
  {\bibfnamefont {Z.}~\bibnamefont {Xia}}, \bibinfo {author} {\bibfnamefont
  {Z.}~\bibnamefont {Ouyang}}, \bibinfo {author} {\bibfnamefont
  {J.}~\bibnamefont {Zhao}}, \bibinfo {author} {\bibfnamefont {H.}~\bibnamefont
  {Guo}},\ and\ \bibinfo {author} {\bibfnamefont {Z.}~\bibnamefont {Tian}},\
  }\bibfield  {title} {\bibinfo {title} {{Large magnetocaloric effect in the
  Shastry-Sutherland lattice compound
  $\mathrm{Y}{\mathrm{b}}_{2}\mathrm{B}{\mathrm{e}}_{2}\mathrm{Ge}{\mathrm{O}}_{7}$
  with spin-disordered ground state}},\ }\href
  {https://doi.org/10.1103/PhysRevB.110.144445} {\bibfield  {journal} {\bibinfo
   {journal} {Phys. Rev. B}\ }\textbf {\bibinfo {volume} {110}},\ \bibinfo
  {pages} {144445} (\bibinfo {year} {2024})}\BibitemShut {NoStop}%
\bibitem [{\citenamefont {Zhang}\ \emph {et~al.}(2026)\citenamefont {Zhang},
  \citenamefont {Na}, \citenamefont {Liu}, \citenamefont {Xiang}, \citenamefont
  {Chen}, \citenamefont {Li}, \citenamefont {Sun}, \citenamefont {Zhou},
  \citenamefont {Zhang},\ and\ \citenamefont {Li}}]{zhang2026refrigeration}%
  \BibitemOpen
  \bibfield  {author} {\bibinfo {author} {\bibfnamefont {Y.}~\bibnamefont
  {Zhang}}, \bibinfo {author} {\bibfnamefont {Y.}~\bibnamefont {Na}}, \bibinfo
  {author} {\bibfnamefont {X.}~\bibnamefont {Liu}}, \bibinfo {author}
  {\bibfnamefont {J.}~\bibnamefont {Xiang}}, \bibinfo {author} {\bibfnamefont
  {F.}~\bibnamefont {Chen}}, \bibinfo {author} {\bibfnamefont {H.-F.}\
  \bibnamefont {Li}}, \bibinfo {author} {\bibfnamefont {P.}~\bibnamefont
  {Sun}}, \bibinfo {author} {\bibfnamefont {S.}~\bibnamefont {Zhou}}, \bibinfo
  {author} {\bibfnamefont {X.}~\bibnamefont {Zhang}},\ and\ \bibinfo {author}
  {\bibfnamefont {L.}~\bibnamefont {Li}},\ }\bibfield  {title} {\bibinfo
  {title} {{Refrigeration down to 0.16 K using a frustrated magnet
  Gd$_2$B$_2$MoO$_9$}},\ }\href {https://doi.org/10.1038/s41467-025-68278-z}
  {\bibfield  {journal} {\bibinfo  {journal} {Nat. Commun.}\ }\textbf {\bibinfo
  {volume} {17}},\ \bibinfo {pages} {1554} (\bibinfo {year}
  {2026})}\BibitemShut {NoStop}%
\bibitem [{\citenamefont {Telang}\ \emph {et~al.}(2025)\citenamefont {Telang},
  \citenamefont {Treu}, \citenamefont {Klinger}, \citenamefont {Tsirlin},
  \citenamefont {Gegenwart},\ and\ \citenamefont
  {Jesche}}]{PhysRevB.111.064431}%
  \BibitemOpen
  \bibfield  {author} {\bibinfo {author} {\bibfnamefont {P.}~\bibnamefont
  {Telang}}, \bibinfo {author} {\bibfnamefont {T.}~\bibnamefont {Treu}},
  \bibinfo {author} {\bibfnamefont {M.}~\bibnamefont {Klinger}}, \bibinfo
  {author} {\bibfnamefont {A.~A.}\ \bibnamefont {Tsirlin}}, \bibinfo {author}
  {\bibfnamefont {P.}~\bibnamefont {Gegenwart}},\ and\ \bibinfo {author}
  {\bibfnamefont {A.}~\bibnamefont {Jesche}},\ }\bibfield  {title} {\bibinfo
  {title} {{Adiabatic demagnetization refrigeration with antiferromagnetically
  ordered ${\mathrm{NaGdP}}_{2}{\mathrm{O}}_{7}$}},\ }\href
  {https://doi.org/10.1103/PhysRevB.111.064431} {\bibfield  {journal} {\bibinfo
   {journal} {Phys. Rev. B}\ }\textbf {\bibinfo {volume} {111}},\ \bibinfo
  {pages} {064431} (\bibinfo {year} {2025})}\BibitemShut {NoStop}%
\bibitem [{\citenamefont {Wang}\ \emph {et~al.}(2024)\citenamefont {Wang},
  \citenamefont {Liu}, \citenamefont {Hu}, \citenamefont {Wang}, \citenamefont
  {Xiang}, \citenamefont {Sun}, \citenamefont {Wang}, \citenamefont {Sun},
  \citenamefont {Zhao}, \citenamefont {Mo} \emph {et~al.}}]{wang2024record}%
  \BibitemOpen
  \bibfield  {author} {\bibinfo {author} {\bibfnamefont {B.}~\bibnamefont
  {Wang}}, \bibinfo {author} {\bibfnamefont {X.}~\bibnamefont {Liu}}, \bibinfo
  {author} {\bibfnamefont {F.}~\bibnamefont {Hu}}, \bibinfo {author}
  {\bibfnamefont {J.-t.}\ \bibnamefont {Wang}}, \bibinfo {author}
  {\bibfnamefont {J.}~\bibnamefont {Xiang}}, \bibinfo {author} {\bibfnamefont
  {P.}~\bibnamefont {Sun}}, \bibinfo {author} {\bibfnamefont {J.}~\bibnamefont
  {Wang}}, \bibinfo {author} {\bibfnamefont {J.}~\bibnamefont {Sun}}, \bibinfo
  {author} {\bibfnamefont {T.}~\bibnamefont {Zhao}}, \bibinfo {author}
  {\bibfnamefont {Z.}~\bibnamefont {Mo}}, \emph {et~al.},\ }\bibfield  {title}
  {\bibinfo {title} {{A record-high cryogenic magnetocaloric effect discovered
  in EuCl$_2$ compound}},\ }\href
  {https://pubs.acs.org/doi/abs/10.1021/jacs.4c12441} {\bibfield  {journal}
  {\bibinfo  {journal} {J. Am. Chem. Soc.}\ }\textbf {\bibinfo {volume}
  {146}},\ \bibinfo {pages} {35016} (\bibinfo {year} {2024})}\BibitemShut
  {NoStop}%
\bibitem [{\citenamefont {Arjun}\ \emph
  {et~al.}(2023{\natexlab{b}})\citenamefont {Arjun}, \citenamefont {Ranjith},
  \citenamefont {Jesche}, \citenamefont {Hirschberger}, \citenamefont {Sarma},\
  and\ \citenamefont {Gegenwart}}]{arjun2023efficient}%
  \BibitemOpen
  \bibfield  {author} {\bibinfo {author} {\bibfnamefont {U.}~\bibnamefont
  {Arjun}}, \bibinfo {author} {\bibfnamefont {K.}~\bibnamefont {Ranjith}},
  \bibinfo {author} {\bibfnamefont {A.}~\bibnamefont {Jesche}}, \bibinfo
  {author} {\bibfnamefont {F.}~\bibnamefont {Hirschberger}}, \bibinfo {author}
  {\bibfnamefont {D.}~\bibnamefont {Sarma}},\ and\ \bibinfo {author}
  {\bibfnamefont {P.}~\bibnamefont {Gegenwart}},\ }\bibfield  {title} {\bibinfo
  {title} {{Efficient adiabatic demagnetization refrigeration to below 50 mK
  with ultrahigh-vacuum-compatible ytterbium diphosphates
  \textit{A}YbP$_2$O$_7$ ($A$ = Na, K)}},\ }\href
  {https://journals.aps.org/prapplied/abstract/10.1103/PhysRevApplied.20.014013}
  {\bibfield  {journal} {\bibinfo  {journal} {Phys. Rev. Appl.}\ }\textbf
  {\bibinfo {volume} {20}},\ \bibinfo {pages} {014013} (\bibinfo {year}
  {2023}{\natexlab{b}})}\BibitemShut {NoStop}%
\bibitem [{\citenamefont {Shimura}\ \emph {et~al.}(2022)\citenamefont
  {Shimura}, \citenamefont {Watanabe}, \citenamefont {Taniguchi}, \citenamefont
  {Osato}, \citenamefont {Yamamoto}, \citenamefont {Kusanose}, \citenamefont
  {Umeo}, \citenamefont {Fujita}, \citenamefont {Onimaru},\ and\ \citenamefont
  {Takabatake}}]{shimura2022magnetic}%
  \BibitemOpen
  \bibfield  {author} {\bibinfo {author} {\bibfnamefont {Y.}~\bibnamefont
  {Shimura}}, \bibinfo {author} {\bibfnamefont {K.}~\bibnamefont {Watanabe}},
  \bibinfo {author} {\bibfnamefont {T.}~\bibnamefont {Taniguchi}}, \bibinfo
  {author} {\bibfnamefont {K.}~\bibnamefont {Osato}}, \bibinfo {author}
  {\bibfnamefont {R.}~\bibnamefont {Yamamoto}}, \bibinfo {author}
  {\bibfnamefont {Y.}~\bibnamefont {Kusanose}}, \bibinfo {author}
  {\bibfnamefont {K.}~\bibnamefont {Umeo}}, \bibinfo {author} {\bibfnamefont
  {M.}~\bibnamefont {Fujita}}, \bibinfo {author} {\bibfnamefont
  {T.}~\bibnamefont {Onimaru}},\ and\ \bibinfo {author} {\bibfnamefont
  {T.}~\bibnamefont {Takabatake}},\ }\bibfield  {title} {\bibinfo {title}
  {{Magnetic refrigeration down to 0.2 K by heavy fermion metal YbCu$_4$Ni}},\
  }\href {https://pubs.aip.org/aip/jap/article/131/1/013903/2836350} {\bibfield
   {journal} {\bibinfo  {journal} {J. Appl. Phys.}\ }\textbf {\bibinfo {volume}
  {131}},\ \bibinfo {pages} {013903} (\bibinfo {year} {2022})}\BibitemShut
  {NoStop}%
\bibitem [{\citenamefont {Jang}\ \emph {et~al.}(2015)\citenamefont {Jang},
  \citenamefont {Gruner}, \citenamefont {Steppke}, \citenamefont {Mitsumoto},
  \citenamefont {Geibel},\ and\ \citenamefont {Brando}}]{jang2015large}%
  \BibitemOpen
  \bibfield  {author} {\bibinfo {author} {\bibfnamefont {D.}~\bibnamefont
  {Jang}}, \bibinfo {author} {\bibfnamefont {T.}~\bibnamefont {Gruner}},
  \bibinfo {author} {\bibfnamefont {A.}~\bibnamefont {Steppke}}, \bibinfo
  {author} {\bibfnamefont {K.}~\bibnamefont {Mitsumoto}}, \bibinfo {author}
  {\bibfnamefont {C.}~\bibnamefont {Geibel}},\ and\ \bibinfo {author}
  {\bibfnamefont {M.}~\bibnamefont {Brando}},\ }\bibfield  {title} {\bibinfo
  {title} {{Large magnetocaloric effect and adiabatic demagnetization
  refrigeration with YbPt$_2$Sn}},\ }\href
  {https://www.nature.com/articles/ncomms9680} {\bibfield  {journal} {\bibinfo
  {journal} {Nat. Commun.}\ }\textbf {\bibinfo {volume} {6}},\ \bibinfo {pages}
  {8680} (\bibinfo {year} {2015})}\BibitemShut {NoStop}%
\bibitem [{\citenamefont {Zhang}\ \emph {et~al.}(2025)\citenamefont {Zhang},
  \citenamefont {Zhang}, \citenamefont {Liu}, \citenamefont {Zhuang},
  \citenamefont {Leng}, \citenamefont {Zhang}, \citenamefont {Xiang},\ and\
  \citenamefont {Sun}}]{zhang2025sub}%
  \BibitemOpen
  \bibfield  {author} {\bibinfo {author} {\bibfnamefont {X.}~\bibnamefont
  {Zhang}}, \bibinfo {author} {\bibfnamefont {T.}~\bibnamefont {Zhang}},
  \bibinfo {author} {\bibfnamefont {X.}~\bibnamefont {Liu}}, \bibinfo {author}
  {\bibfnamefont {Z.}~\bibnamefont {Zhuang}}, \bibinfo {author} {\bibfnamefont
  {Z.}~\bibnamefont {Leng}}, \bibinfo {author} {\bibfnamefont {S.}~\bibnamefont
  {Zhang}}, \bibinfo {author} {\bibfnamefont {J.}~\bibnamefont {Xiang}},\ and\
  \bibinfo {author} {\bibfnamefont {P.}~\bibnamefont {Sun}},\ }\bibfield
  {title} {\bibinfo {title} {{Sub-Kelvin magnetocaloric effect in frustrated
  intermetallic NdNi$_4$Mg}},\ }\href
  {https://pubs.aip.org/aip/jap/article/138/6/063903/3358331/Sub-Kelvin-magnetocaloric-effect-in-frustrated}
  {\bibfield  {journal} {\bibinfo  {journal} {J. Appl. Phys.}\ }\textbf
  {\bibinfo {volume} {138}},\ \bibinfo {pages} {063903} (\bibinfo {year}
  {2025})}\BibitemShut {NoStop}%
\end{thebibliography}%
%

\clearpage
\newpage
\renewcommand{\thefigure}{S\arabic{figure}}
\renewcommand{\thetable}{S\arabic{table}}
\setcounter{figure}{0}
\setcounter{table}{0}
\setcounter{section}{0}

\onecolumngrid
\begin{center}
    {\large \bfseries Supporting Information for ``Giant magnetocaloric effect at low fields in triangular-lattice \nmao''}
\end{center}
\vspace{1cm}


\section{Single-crystal XRD}
\begin{figure*}[!h]
	\centering
	\includegraphics[width=0.55\columnwidth]{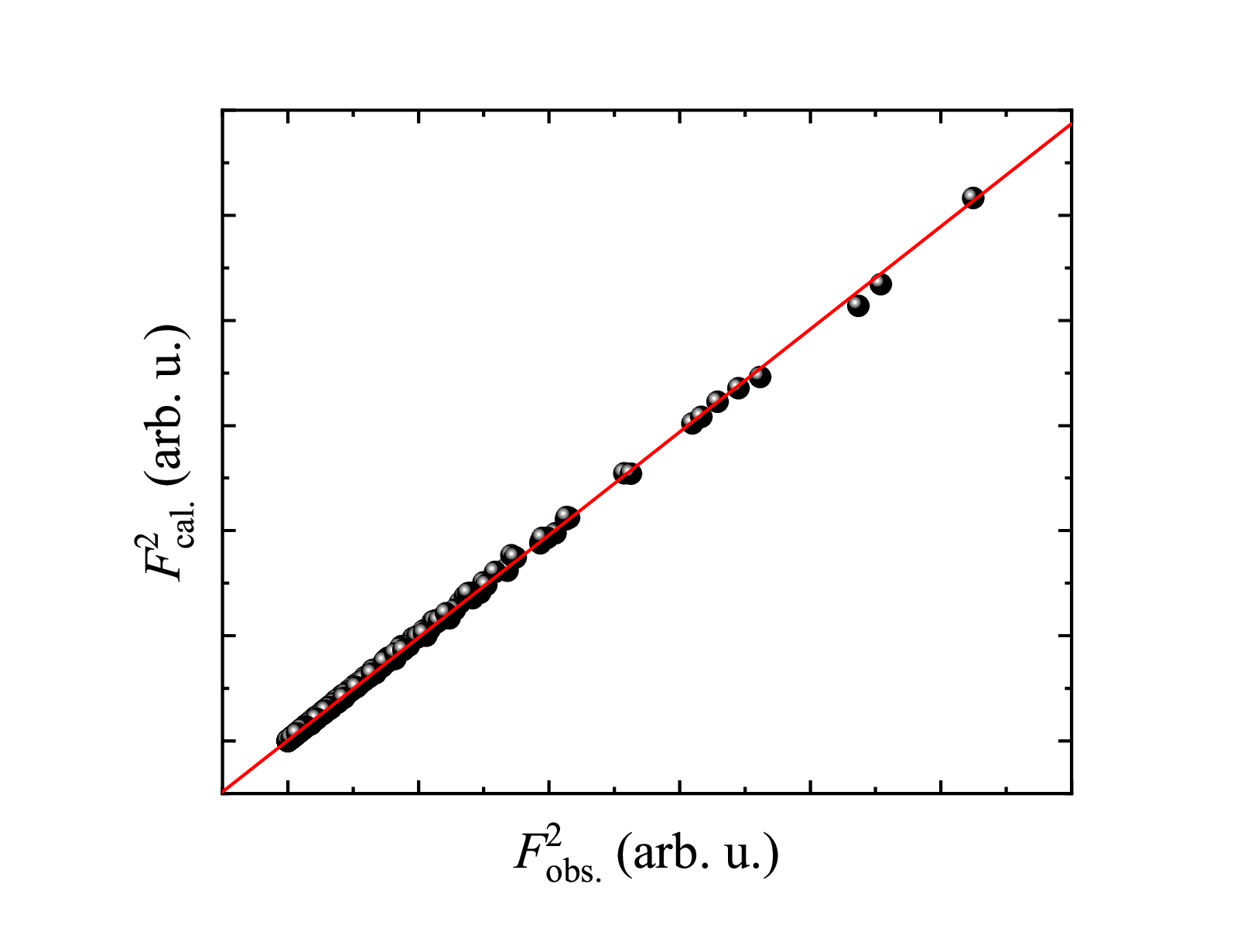}
	\caption{$F^2_{obs.}$ vs. $F^2_{cal.}$ from single-crystal structural refinement. The straight line is a guide to the eye.}
	\label{sfig1}
\end{figure*}

Single-crystal X-ray diffraction conditions are listed in Tab. \ref{stab1}. The refined crystal structure parameters are tabulated in Tab. \ref{stab2}. For a direct visualization of the refinement quality, the $F^2_\mathrm{obs.}$ against $F^2_\mathrm{cal.}$ is shown in Fig. \ref{sfig1}.

\begin{table}[!h]
	\caption{Experimental conditions for the single crystal XRD measurements, and agreement factors for the refinement.\label{stab1}}
\centering
	\begin{tabular}{l r}
		\hline
		\hline
		Formula                                                      & NdMgAl$_{11}$O$_{19}$ \\
		Space group                                                  & $P6_3/mmc$ (No. 194) \\
		\textit{a, b} ($\textrm{\AA}$)                               & 5.57549(19) \\
		\textit{c} ($\textrm{\AA}$)                                  & 21.8733(8) \\
		\textit{V} ($\textrm{\AA}^3$)                                & 588.86(4) \\
		\textit{Z}                                                   & 2 \\
		2$\Theta$ ($^\circ$)                                         & 7.46 - 81.46 \\
		No. of measured reflections, $R_\mathrm{int}$               & 27315, 4.92\% \\
		No. of independent reflections                               & 718 \\
		No. of parameters                                            & 47 \\
		\multirow{3}*{Index ranges}                                  & -10 $\leq$ \textit{H} $\leq$ 10, \\
		& -10 $\leq$ \textit{K} $\leq$ 10, \\
		& -39 $\leq$ \textit{L} $\leq$ 39 \\
		\textit{R}, \textit{wR$_2$}                    & 1.26\%, 3.09\% \\
		Goodness of fit on $F^2$                                     & 1.13 \\
		Largest difference peak/hole (\textit{e}/$\textrm{\AA}^{3}$)& 0.19 / -0.57 \\
		\hline
		\hline
	\end{tabular}
\end{table}

\begin{table}
	\caption{Refined crystal structure parameters.\label{stab2}}
	\begin{tabular}{ccccc}
		\hline
		\hline
		Atom  & occ. & x & y & z \\
		
		Nd1 (2d)  & 0.9158(17) & 1/3 & 2/3 & 0.75 \\
		Nd2 (6h)  & 0.0281(6)  & 0.2641(5) & 0.7359(5) & 0.75 \\
		Al1 (2a)  & 1          & 0 & 0 & 0.5 \\
		Al2 (12k) & 1          & 0.66491(4) & 0.832453(18) & 0.608500(11) \\
		Al3 (4f)  & 1          & 2/3 & 1/3 & 0.690041(19) \\
		Mg (4f)  & 0.5        & 2/3 & 1/3 & 0.527285(19) \\
		Al4 (4f)  & 0.5        & 2/3 & 1/3 & 0.527285(19) \\
		Al5 (4e)  & 0.5        & 0 & 0 & 0.74245(8) \\
		O1 (4e)   & 1          & 0 & 0 & 0.65159(4) \\
		O2 (6h)   & 1          & 0.36264(15) & 0.18132(8) & 0.75 \\
		O3 (12k)  & 1          & 0.49453(5) & 0.50547(5) & 0.65175(2) \\
		O4 (12k)  & 1          & 0.84763(5) & 0.69525(11) & 0.55367(3) \\
		O5 (4f)   & 1          & 1/3 & 2/3 & 0.55795(5) \\
		\hline
		
	\end{tabular}
	
	\begin{tabular}{cccc}
		Atom & $U_{11}/U_{12}$ & $U_{22}/U_{13}$ & $U_{33}/U_{23}$ \\
		Nd1 & 0.01118(6) & 0.01118(6) &  0.00652(5) \\
		&0.00559(3)&0&0\\
		Nd2 & 0.0184(9)  &0.0184(9)  & 0.0137(9)  \\
		& 0.0013(9) & 0 & 0 \\
		Al1 & 0.00368(11) &0.00368(11) & 0.00330(17) \\
		&0.00184(5) & 0 & 0 \\
		Al2 & 0.00358(9)  &0.00356(7)  & 0.00401(9)  \\
		& 0.00179(5) & 0.00001(4) & 0.00000(2) \\
		Al3 & 0.00380(9)  &0.00380(9)  & 0.00303(13) \\
		& 0.00190(5) & 0 & 0 \\
		Mg/Al4 & 0.00317(10) &0.00317(10) & 0.00359(16) \\
		& 0.00158(5) & 0 & 0 \\
		Al5 & 0.00470(14) & 0.00470(14) & 0.0153(11) \\
		& 0.00235(7) & 0 & 0 \\
		O1 & 0.00429(16) &0.00429(16) & 0.0071(3) \\
		& 0.00214(8) & 0 & 0 \\
		O2 & 0.0046(2) & 0.0112(2) &0.0043(2) \\
		& 0.00231(11) & 0 & 0 \\
		O3 & 0.00416(12) &0.00416(12) & 0.00566(18) \\
		& 0.00193(12) & -0.00078(6) & 0.00078(6) \\
		O4 & 0.00619(13) & 0.00869(17)&0.00556(19) \\
		& 0.00435(9) & 0.00098(7) & 0.00196(13) \\
		O5 & 0.00442(18) &0.00442(18) & 0.0068(3) \\
		& 0.00221(9) & 0 & 0 \\
		\hline
		\hline
	\end{tabular}
\end{table}

\section{Powder XRD}

\begin{figure*}
	\centering
	\includegraphics[width=0.65\columnwidth]{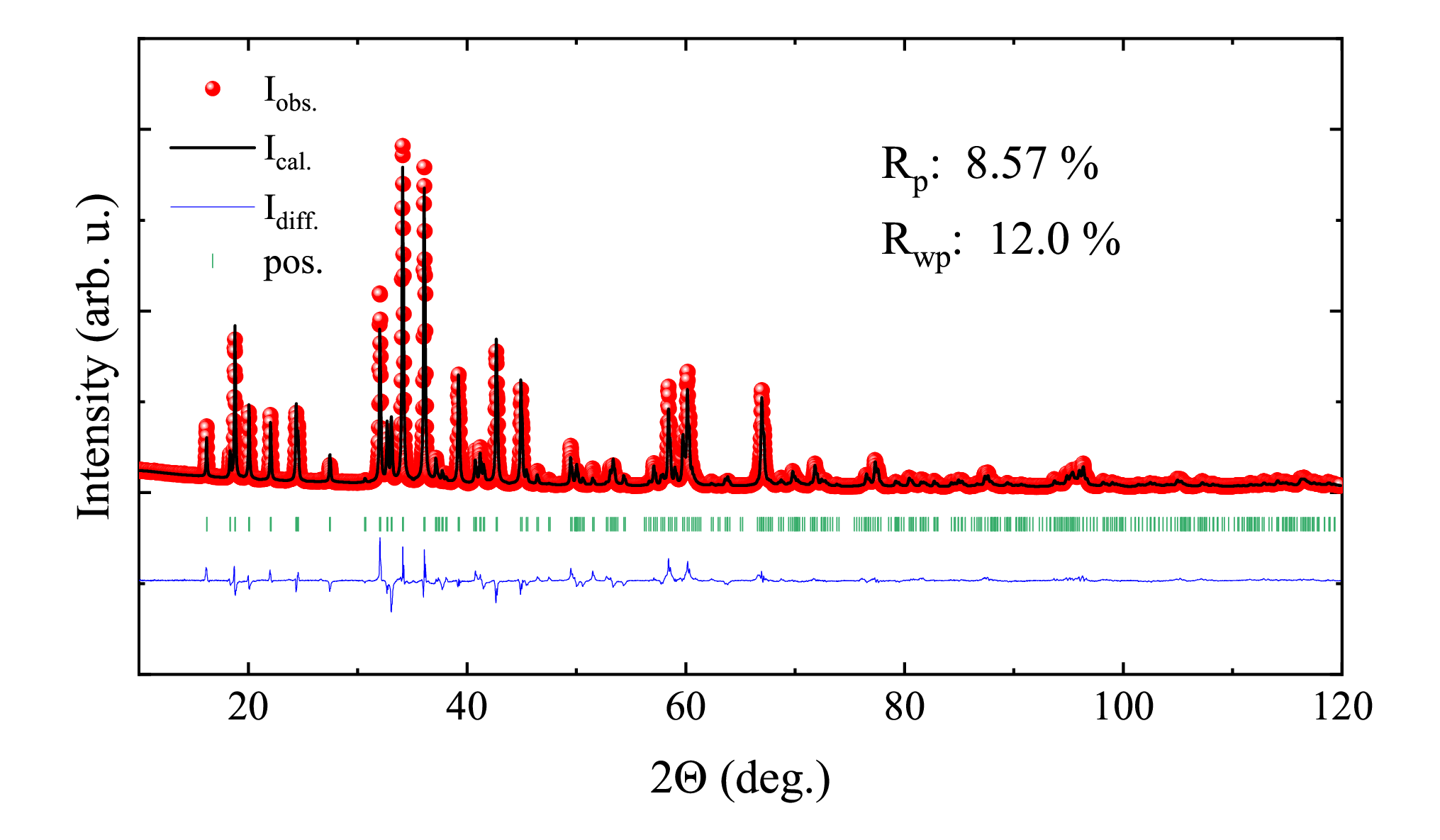}
	\caption{Rietveld structural refinement against the powder X-ray diffraction pattern for \nmao.}
	\label{sfig2}
\end{figure*}

The powder X-ray diffraction pattern for the crushed single crystals is shown in Fig. \ref{sfig2}. Due to the strong preferred orientation related to the layered structure of the sample, the final agreement factors are relatively high even when the preferred orientation was considered during the refinement using the Fullprof package. However, the key information is that no additional peaks were observed from this measurement, indicating an impurity-free phase within our experimental resolution.

\end{document}